\documentclass[12pt]{article}
\usepackage{epsfig}
\usepackage{amssymb,amsmath}
\usepackage{setspace}
\pdfoutput=1
\usepackage{subfigure}
\usepackage{amssymb,amsmath}
\usepackage{graphicx}
\usepackage{color}
\usepackage{cancel}
\usepackage[colorlinks=true
,urlcolor=blue
,citecolor=blue
,linkcolor=blue
,pagecolor=blue
,linktocpage=true
,pdfproducer=medialab
]{hyperref}
\def\bi{\begin{itemize}}
\def\ei{\end{itemize}}

\def\tl{\tilde l}

\def\tst{\tilde t}
\def\ttau{\tilde \tau}

\def\alt{\lesssim}
\def\agt{\gtrsim}

\newcommand\prd[3]{{\it Phys.\ Rev.\ }{\bf D #1} (#2) #3}
\newcommand\prl[3]{{\it Phys.\ Rev.\ Lett.\ }{\bf #1} (#2) #3}

\newcommand\plb[3]{{\it Phys.\ Lett.\ }{\bf B #1} (#2) #3}

\newcommand\jhep[3]{{\it J. High Energy Phys.\ }{\bf #1} (#2) #3}

\def\tl{\tilde l}

\def\tst{\tilde t}
\def\ttau{\tilde \tau}

\def\tl{\tilde}

\usepackage{amsmath}

\newcommand{\bea}{\begin{eqnarray}}
\newcommand{\eea}{\end{eqnarray}}

\newcommand{\beq}{\begin{equation}}
\newcommand{\eeq}{\end{equation}}

 \makeatletter
\def\alt{\mathrel{\mathpalette\gl@align<}}
\def\agt{\mathrel{\mathpalette\gl@align>}}
\def\gl@align#1#2{\lower.6ex\vbox{\baselineskip\z@skip\lineskip\z@
\ialign{$\m@th#1\hfil##\hfil$\crcr#2\crcr\sim\crcr}}} \makeatother

\usepackage{longtable}

\begin{document}
%
\vspace*{1.0cm}

\begin{center}
\baselineskip 20pt {\Large\bf
The Electroweak Supersymmetry (EWSUSY) from the GmSUGRA in the MSSM
}
\vspace{1cm}

{\large
Tianjun Li$^{a,b}$\footnote{E-mail:tli@itp.ac.cn},
Shabbar Raza$^{a}$\footnote{E-mail:shabbar@itp.ac.cn}, 
} \vspace{.5cm}

{\baselineskip 20pt \it $^a$
State Key Laboratory of Theoretical Physics and Kavli Institute for Theoretical Physics China (KITPC),
Institute of Theoretical Physics, Chinese Academy of Sciences, Beijing 100190, P. R. China \\
}
{\it $^b$
School of Physical Electronics, University of Electronic Science and Technology of China,\\
Chengdu 610054, P. R. China \\
}

\vspace{.5cm}

\vspace{1.25cm} {\bf Abstract}
\end{center}

Considering the Generalized Minimal Supergravity Model (GmSUGRA) in the Minimal Supersymmetric 
Standard Model (MSSM), we study the Electroweak Supersymmetry (EWSUSY), where the squarks and/or 
gluino are heavy around a few TeVs while the sleptons, sneutrinos, bino, winos, and/or higgsinos 
are light within one TeV. We resolves the $(g-2)_{\mu}/2$ discrepancy for the muon anomalous 
magnetic moment in the Standard Model (SM) 
successfully and identifies a parameter space where such solutions also have the electroweak 
fine-tuning measures $\Delta_{EW}~16.5$ (6$\%$) and $\Delta_{EW}~25$ (4$\%$) without and with 
the Wilkinson Microwave Anisotropy Probe (WMAP) bounds, respectively. We find that the allowed mass ranges, which are consistent within 3$\sigma$ 
of the $g-2$ discrepancy, for the lightest neutralino, charginos, stau, stau neutrinos, 
and firse two families of sleptons are $[44,390]$ GeV, $[100,700]$ GeV, $[100,700]$, $[52,800]$ and $[150,800]$ GeV, 
respectively. Moreover, our solutions satisfy the latest bounds reported by the ATLAS and CMS 
Collaborations on electroweakinos and sleptons. The colored sparticles such as light stop, gluinos, 
and the first two generations of squark masses have been found in the mass ranges of $[500, 3000]$ GeV, 
[1300, 4300] GeV, and $[1800, 4200]$ GeV, respectively. To obtain the observed dark matter relic density 
for the Lightest Supersymmetric Particle (LSP) neutralino, we have the bino-wino, LSP neutralino-stau, and 
LSP neutralino-tau sneutrinos coannihilation scenarios, and the resonance solutions such as $A$-pole, 
Higgs-pole, and $Z$-pole. We identify the higgsino-like LSP neutralino and display its spin-independent 
and spin-dependent cross sections with nucleons. We present ten benchmark 
points which can be tested at the up coming collider searches as well.

\thispagestyle{empty}

\newpage

\addtocounter{page}{-1}

\baselineskip 18pt

\section{Introduction}

It is well-known that supersymmetry (SUSY) provides a natural solution to the gauge hierarchy problem
in the Standard Model (SM). In the supersymmetric SMs (SSMs), gauge coupling unification 
can be realized which strongly indicates the Grand Unified Theories (GUTs), and the electroweak (EW) 
gauge symmetry can be broken radiatively due to the large top quark Yukawa coupling. If conservation of $R$-parity
is assumed, the Lightest Supersymmetric Particle (LSP) such as neutralino is a dark matter candidate. Thus, 
SUSY is the most promising new physics beyond the SM.

From the first run of the LHC, a SM-like Higgs boson with mass $m_h$ around 125 GeV was discovered 
in July 2012~\cite{ATLAS, CMS}. This is a little bit heavy for 
the Minimal SSM (MSSM) since it requires the multi-TeV top squarks with small mixing or TeV-scale 
top squarks with large mixing. Moreover, we have strong constraints on the parameter space in the SSMs from the LHC SUSY searches.
For example, the gluino mass $m_{\tilde g}$ should be heavier than about 1.7 TeV if the first 
two-generation squark mass $m_{\tilde q}$ is around the gluino mass $m_{\tilde q} \sim m_{\tilde g}$, 
and heavier than about 1.3 TeV for $m_{\tilde q} \gg m_{\tilde g}$~\cite{Chatrchyan:2013wxa, Aad:2014wea}.

Inspired by the LHC Higgs~\cite{moriond2013} and SUSY~\cite{LHC-SUSY} searches, as well as 
the experimental results/constraints
on B physics~\cite{Aaij:2012nna, Buchmueller:2009fn} 
and Flavor Changing Neutral Current 
(FCNC)~\cite{Barberio:2007cr, Asner:2010qj, Amhis:2012bh}, anomalous
magnetic momentum of the muon~\cite{Davier:2010nc,Bennett:2006fi},  
dark matter relic density from WMAP experiment~\cite{WMAP9}, and 
direct dark matter search from LUX experiment~\cite{Akerib:2013tjd}, 
one of us (TL) with his collaborators proposed the Electroweak Supersymmetry (EWSUSY), where
the squarks and/or gluino are heavy around a few TeVs while the sleptons, sneutrinos, bino, 
winos, and/or higgsinos are light within one TeV~\cite{Cheng:2012np}. Especially,
the EWSUSY can be realized in the
Generalized Minimal Supergravity (GmSUGRA)~\cite{Li:2010xr, Balazs:2010ha}.

In this paper, we shall systematically study the SM $(g-2)_{\mu}/2$ discrepancy for the muon anomalous 
magnetic moment in the MSSM with the EWSUSY from GmSUGRA.
 We find that the EWSUSY from GmSUGRA not only resolves the $(g-2)_{\mu}/2$ anomaly but also addresses the
Electroweak Fine Tuning (EWFT) problem. We show the preferred mass ranges for some SUSY Breaking
(SSB) terms required to explain the muon $(g-2)_{\mu}/2$ anomaly. It is well-known that neutralinos, charginos 
(collectively known as electroweakinos), and sleptons play very 
important roles in addressing the muon $(g-2)_{\mu}/2$ anomaly.  We show that the EWSUSY from 
GmSUGRA very effectively resolves the muon $(g-2)_{\mu}/2$ anomaly. The allowed mass ranges consistent within 
3$\sigma$ of $(g-2)_{\mu}/2$ discrepancy for the LSP neutralino, charginos, stau, stau neutrinos,
 and first two families of sleptons are $[44,390]$ GeV, $[100,700]$ GeV, $[100,700]$, and $[52,800]$ and $[150,800]$ GeV,
 respectively. Recently, the ATLAS and CMS Collaborations
have reported new bounds on electroweakinos as well as 
all three families of sleptons and sneutrinos depending on various assumptions and
topologies. We discuss these bounds in some detail and find that our solutions are consistent with these bounds and
still provide resolution to the muon magnetic dipole moment anomaly within 3$\sigma$. We also note that some portions the parameter
space are not only consistent with all the collider and astrophysical bounds but also provides even within $1 \sigma$ 
contributions to the muon $(g-2)_{\mu}/2$ and hence resolves discrepancy successfully. 
 For color sparticles, we note that the
light stop is the lightest colored sparticle in our data having mass range $[500,3000]$ GeV, while gluino mass range is
$[1300,4300]$ GeV. This gluino mass range shrinks a little to 3000 GeV if we insist on dark matter 
relic density bounds. The first two families of squarks lie in the mass ranges from 
1800 GeV to 4200 GeV. We also identify a viable parameter space 
which satisfies all the bounds including 5$\sigma$ WMAP9 bounds, resolves the muon $(g-2)_{\mu}/2$ anomaly, 
as well as provides
solutions with small EWFT. We note that in our data the minimal EWFT measures $\Delta_{EW}\sim 16.5$ (6$\%$) 
and $\Delta_{EW}\sim 25$ (4$\%$) without and with the WMAP9 
bound, respectively. In our present scans we find that in order to obtain the observed
dark matter relic density, we have the bino-wino, LSP neutralino-stau, LSP neutralino-tau sneutrino 
coannihilation scenarios and resonance
solutions such as $A$-resonance, Higgs-resonance and $Z$-resonance for bino-like neutralino. 
Moreover, we comment on the bino-like
solutions which do not satisfy the WMAP9 bounds. Apart from the bino-like LSP, we have wino-like and higgsino-like
LSPs. These  wino-like and higgsino-like
LSPs solutions have very small relic density. We comment on such  wino-like LSP solutions. We display graphs 
for direct and indirect searches for higgsino-like LSP.
Finally we present ten benchmark points in two tables showing some characteristic features of our models.

This paper is organized as follows. In Section~\ref{section_2}, we briefly describe the GmSUGRA model 
and the SSB parameters.
We also briefly discuss $(g_{\mu}-2)/2$ anomaly and describe our definition of EWFT. 
In Section~\ref{sec:scan}, we outline the detailed
SSB parameters, the ranges of numerical values employed in our scan, the scanning procedure,
and the relevant experimental constraints that we have considered. 
We discuss results of our scans in Section~\ref{results}. 
A summary and conclusion are given in Section \ref{summary}.

\section{The EWSUSY from the GmSUGRA in the MSSM}
\label{section_2}

In the GmSUGRA~\cite{Li:2010xr, Balazs:2010ha}, one can realize the EWSUSY, where the sleptons and electroweakinos 
(charginos, bino, wino, and/or higgsinos) are within one TeV while squarks and/or gluinos 
can be in several TeV mass ranges~\cite{Cheng:2012np}. Moreover, the gauge
coupling relation and gaugino mass relation at the GUT scale are 
\begin{equation}
 \frac{1}{\alpha_2}-\frac{1}{\alpha_3} =
 k~\left(\frac{1}{\alpha_1} - \frac{1}{\alpha_3}\right)~,
\end{equation}
\begin{equation}
 \frac{M_2}{\alpha_2}-\frac{M_3}{\alpha_3} =
 k~\left(\frac{M_1}{\alpha_1} - \frac{M_3}{\alpha_3}\right)~,
\end{equation}
where $k$ is the index and equal to 5/3 in the simple GmSUGRA. We obtain a simple gaugino mass relation
\begin{equation}
 M_2-M_3 = \frac{5}{3}~(M_1-M_3)~,
\label{M3a}
\end{equation}
by assuming gauge coupling unification at the GUT scale ($\alpha_1=\alpha_2=\alpha_3$). It is obvious that
the universal gaugino mass relation $M_1 = M_2 = M_3$ in the mSUGRA, is just a special case of 
this general one. Choosing $M_1$ and $M_2$ to be free input parameters, which vary around 
several hundred GeV for the EWSUSY, we get $M_3$ from Eq.~(\ref{M3a}):
\begin{eqnarray}
M_3=\frac{5}{2}~M_1-\frac{3}{2}~M_2~,
\label{M3}
\end{eqnarray}
which could be as large as several TeV or as small as several hundred GeV, depending
 on specific values of $M_1$ and $M_2$.

The general SSB scalar masses at the GUT scale are given 
in Ref.~\cite{Balazs:2010ha}. 
Taking the slepton masses as free parameters, we obtain the following squark masses 
in the $SU(5)$ model with an adjoint Higgs field
\begin{eqnarray}
m_{\tl{Q}_i}^2 &=& \frac{5}{6} (m_0^{U})^2 +  \frac{1}{6} m_{\tl{E}_i^c}^2~,\\
m_{\tl{U}_i^c}^2 &=& \frac{5}{3}(m_0^{U})^2 -\frac{2}{3} m_{\tl{E}_i^c}^2~,\\
m_{\tl{D}_i^c}^2 &=& \frac{5}{3}(m_0^{U})^2 -\frac{2}{3} m_{\tl{L}_i}^2~,
\label{squarks_masses}
\end{eqnarray}
where $m_{\tl Q}$, $m_{\tl U^c}$, $m_{\tl D^c}$, $m_{\tl L}$, and  $m_{\tl E^c}$ represent the scalar masses of
the left-handed squark doublets, right-handed up-type squarks, right-handed down-type squarks,
left-handed sleptons, and right-handed sleptons, respectively, while $m_0^U$ is the universal  
scalar mass, as in the mSUGRA. In the EWSUSY, $m_{\tl L}$ and $m_{\tl E^c}$ are both within 1 TeV, resulting in 
light sleptons. Especially, in the limit $m_0^U \gg m_{\tl L/\tl E^c}$, we have the approximated 
relations for squark masses: $2 m_{\tl Q}^2 \sim m_{\tl U^c}^2 \sim m_{\tl D^c}^2$. In addition, 
the Higgs soft masses $m_{\tl H_u}$ and $m_{\tl H_d}$, and the  trilinear soft terms
 $A_U$, $A_D$ and $A_E$ can all be free parameters from the GmSUGRA~\cite{Cheng:2012np, Balazs:2010ha}.

\subsection{The anomalous magnetic moment of the muon
 $a_{\mu}=(g-2)_{\mu}/2$}
\label{g-2}
In parallel to the on-going searches for the new physics at the high energy collider, one can look for such
effects at low energy. The precise measurement of muon $a_{\mu}=(g-2)_{\mu}/2$
 may reveal, though indirectly, traces for the 
physics beyond the SM. The SM prediction for the anomalous magnetic moment 
of the muon~\cite{Hagiwara:2011af} shows a discrepancy 
with the experimental results~\cite{Bennett:2006fi}, 
which is quantified as follows
\begin{eqnarray}
\Delta a_{\mu}\equiv a_{\mu}({\rm exp})-a_{\mu}({\rm SM})= (28.6 \pm 8.0) \times 10^{-10}~.~\,
\label{bound2}
\end{eqnarray}
If SUSY does exist at the EW scale, then the main SUSY
contributions to $a_{\mu}$ come from the neutralino-smuon and chargino-sneutrino loops 
and are given as~\footnote{For complete one-loop result, see Ref.~\cite{Moroi:1995yh}.}
\begin{eqnarray}
\Delta a_{\mu}^{SUSY} \propto \frac{M_{i}\mu \tan\beta}{m_{SUSY}^{4}}~,
\label{amu_susy}
\end{eqnarray}
where $M_{i}$($i=1,2$) are the weak scale gaugino masses, $\mu$ is the higgsino mass parameter, 
$\tan\beta\equiv\frac{\langle H_u \rangle}{\langle H_d\rangle}$,
and $m_{SUSY}$ is the sparticle mass circulating in the loop. It is also evident from Eq.~(\ref{amu_susy})
that by having appropriately light $m_{SUSY}$ masses (electroweakinos and sleptons), we may have
sizable SUSY contributions to $\Delta a_{\mu}$. In order to address the $g-2$ anomaly between experiment and theory, new direct
measurements of the muon magnetic moment with fourfold improvement in accuracy have been proposed at Fermilab by E989 experiment, and
 Japan Proton Accelerator Research Complex~\cite{Venanzoni:2014ixa}. First results from E989 are expected around 2017/18. 
These measurements will firmly establish or constrain new physics effects. Spurred by these developments new studies have been done in order 
to explore this opportunity~\cite{Cheng:2012np,Badziak:2014kea}.  In this article while doing general scans we resolve 
the muon $(g-2)_{\mu}/2$ successfully and add new dark matter channels such as Higgs-resonance and $Z$-resonance consistent with $\Delta a_{\mu}$ values within 3$\sigma$ in addition to the previously reported channels~\cite{Cheng:2012np,Gogoladze:2014cha}. Moreover, we show that our solutions while having previously
mentioned properties, also have small electroweak fine-tuning (defined below). In our scans,
the sleptons and electroweakinons mass ranges, which are required to address the $(g-2)_{\mu}/2$ problem, are in agreement with 
Refs.~\cite{Cheng:2012np,Gogoladze:2014cha,Endo:2013bba}.


\subsection{The Electroweak Fine Tuning}\label{ewft}

It is interesting to note that in addition to resolve $a_{\mu}$ anomaly, the EWSUSY from 
GmSUGRA can also accommodate the solutions
with small EWFT. In the first site it appears contradictory. At one hand, from 
Eq.~(\ref{amu_susy}) it appears that the large values of $\mu$ are required for sizable $a_{\mu}^{SUSY}$ contributions. 
On the other hand, small EWFT requires small values of $\mu$. But after looking at Eq.~(\ref{amu_susy})
more carefully, we see that by having suitable large values for gaugino masses and $\tan\beta$, 
and small values for electroweakino and slepton masses, one can compensate the small values of $\mu$ 
(required for small EWFT) and still resolve $a_{\mu}$ anomaly.   

We use the latest (7.84) version of  ISAJET \cite{ISAJET} to calculate the  fine-tuning (FT) conditions 
at the EW scale $M_{EW}$. After including the one-loop effective potential contributions to the tree-level MSSM 
Higgs potential, the $Z$-bosom mass $M_Z$ is given by
\begin{equation}
\frac{M_Z^2}{2} =
\frac{(m_{H_d}^2+\Sigma_d^d)-(m_{H_u}^2+\Sigma_u^u)\tan^2\beta}{\tan^2\beta
-1} -\mu^2 \; ,
\label{eq:mssmmu}
\end{equation}
where $\Sigma_u^u$ and  $\Sigma_d^d$ are the contributions coming from the one-loop effective potential 
defined in Ref.~\cite{Baer:2012mv} and $\tan\beta \equiv \frac{v_u}{v_d}$. All parameters  
in Eq. (\ref{eq:mssmmu}) are defined at the $M_{EW}$.
In order to measure the EWFT condition we follow \cite{Baer:2012mv} and use the following definitions
\begin{equation}
 C_{H_d}\equiv |m_{H_d}^2/(\tan^2\beta -1)|,\,\, C_{H_u}\equiv
|-m_{H_u}^2\tan^2\beta /(\tan^2\beta -1)|, \, \, C_\mu\equiv |-\mu^2 |,
\label{cc1}
\end{equation}
 with
each $C_{\Sigma_{u,d}^{u,d} (k)}$  less than some characteristic value of order $M_Z^2$.
Here, $k$ labels the SM and SUSY particles that contribute to the one-loop Higgs potential.
For the fine-tuning measure we define
\begin{equation}
 \Delta_{\rm EW}\equiv {\rm max}(C_k )/(M_Z^2/2)~.
\label{eq:ewft}
\end{equation}
Note that $\Delta_{EW}$ only depends on the weak-scale parameters of the SSMs, and then is fixed
by the particle spectra. Hence, it is independent of how the SUSY particle masses arise. 
Lower values of $\Delta_{EW}$
corresponds to less fine tuning, for example, $\Delta_{EW}=10$ implies $\Delta_{EW}^{-1}=10\%$ fine tuning.
In addition to $\Delta_{EW}$, ISAJET also calculates $\Delta_{HS}$ which is a measure of fine-tuning at 
the High Scale (HS) like the GUT scale
in our case ~\cite{Baer:2012mv}. The HS scale fine-tuning measure $\Delta_{HS}$ is given as follows
\begin{equation}
 \Delta_{\rm HS}\equiv {\rm max}(B_i )/(M_Z^2/2)~.
\label{eq:hsft}
\end{equation}
For definition of $B_i$ and more details see Ref.~\cite{Baer:2012mv}. 

\section{Phenomenological Constraints and Scanning Procedure}
\label{sec:scan}

We employ the ISAJET~7.84 package~\cite{ISAJET}
 to perform random scans over the parameter space
 given below.
In this package, the weak scale values of the gauge and third
 generation Yukawa couplings are evolved to
 $M_{\rm GUT}$ via the MSSM renormalization group equations (RGEs)
 in the $\overline{DR}$ regularization scheme.
We do not strictly enforce the unification condition
 $g_3=g_1=g_2$ at $M_{\rm GUT}$, since a few percent deviation
 from unification can be assigned to the unknown GUT-scale threshold
 corrections~\cite{Hisano:1992jj}.
With the boundary conditions given at $M_{\rm GUT}$,
 all the SSB parameters, along with the gauge and Yukawa couplings,
 are evolved back to the weak scale $M_{\rm Z}$.

In evaluating Yukawa couplings the SUSY threshold
 corrections~\cite{Pierce:1996zz} are taken into account
 at the common scale $M_{\rm SUSY}= \sqrt{m_{\tst_L}m_{\tst_R}}$.
The entire parameter set is iteratively run between
 $M_{\rm Z}$ and $M_{\rm GUT}$ using the full two-loop RGEs
 until a stable solution is obtained.
To better account for the leading-log corrections, one-loop step-beta
 functions are adopted for gauge and Yukawa couplings, and
 the SSB parameters $m_i$ are extracted from RGEs at appropriate scales
 $m_i=m_i(m_i)$.
The RGE-improved one-loop effective potential is minimized
 at an optimized scale $M_{\rm SUSY}$, which effectively
 accounts for the leading two-loop corrections.
The full one-loop radiative corrections are incorporated
 for all sparticles.

The requirement of radiative electroweak symmetry breaking
 (REWSB)~\cite{Ibanez:1982fr} puts an important theoretical
 constraint on parameter space.
Another important constraint comes from limits on the cosmological
 abundance of stable charged particle~\cite{Beringer:1900zz}.
This excludes regions in the parameter space where charged
 SUSY particles, such as $\ttau_1$ or $\tst_1$,
 become the LSP.
We accept only those solutions for which one of the neutralinos
 is the LSP.

Using parameters given in Section~\ref{section_2}, we have performed 
the random scans
 for the following parameter ranges
\begin{align}
100 \, \rm{GeV} \leq & m_0^{U}  \leq 5000 \, \rm{GeV}  ~,~\nonumber \\
100 \, \rm{GeV} \leq & M_1  \leq 900 \, \rm{GeV} ~,~\nonumber \\
100\, \rm{GeV} \leq & M_2   \leq 800 \, \rm{GeV} ~,~\nonumber \\
100 \, \rm{GeV} \leq & m_{\tilde L}  \leq 800 \, \rm{GeV} ~,~\nonumber \\
100 \, \rm{GeV} \leq & m_{\tilde E^c}  \leq 800 \, \rm{GeV} ~,~\nonumber \\
100 \, \rm{GeV} \leq & m_{\tilde H_{u,d}} \leq 5000 \, \rm{GeV} ~,~\nonumber \\
-6000 \, \rm{GeV} \leq & A_{U}=A_{D} \leq 5000 \, \rm{GeV} ~,~\nonumber \\
-800 \, \rm{GeV} \leq & A_{E} \leq 935 \, \rm{GeV} ~,~\nonumber \\
2\leq & \tan\beta  \leq 60~.~
 \label{input_param_range}
\end{align}
Also, we consider  $\mu > 0$ and  use $m_t = 173.3\, {\rm GeV}$  \cite{:2009ec}.
Note that our results are not too sensitive to one
 or two sigma variation in the value of $m_t$  \cite{bartol2}.
We use $m_b^{\overline{DR}}(M_{\rm Z})=2.83$ GeV as well
 which is hard-coded into ISAJET. Also note that we will use the notations $A_t,A_b,A_{\tau}$ for $A_{U},A_D$ and $A_E$ receptively.


In scanning the parameter space, we employ the Metropolis-Hastings
 algorithm as described in \cite{Belanger:2009ti}.
The data points collected all satisfy the requirement of REWSB,
 with the neutralino being the LSP.
After collecting the data, we require the following bounds (inspired by the LEP2 experiment) 
on sparticle masses
\begin{eqnarray} 
m_{\tilde t_1},m_{\tilde b_1} \gtrsim 100 \; {\rm GeV} ~,~\\
m_{\tilde \tau_1} \gtrsim 105 \; {\rm GeV}  ~,~\\
m_{\tilde \chi_{1}^{\pm}} \gtrsim 103 \; {\rm GeV}~.~
\end{eqnarray}

Moreover, we use the IsaTools package~\cite{bsg, bmm} and Ref.~\cite{mamoudi}
 to implement the following B-physics constraints
\begin{eqnarray}
0.8\times 10^{-9} \leq{\rm BR}(B_s \rightarrow \mu^+ \mu^-) 
  \leq 6.2 \times10^{-9} \;(2\sigma)~~&\cite{Aaij:2012nna} ~,~&
\\ 
2.99 \times 10^{-4} \leq 
  {\rm BR}(b \rightarrow s \gamma) 
  \leq 3.87 \times 10^{-4} \; (2\sigma)~~&\cite{Amhis:2012bh}  ~,~&
\\
0.15 \leq \frac{
 {\rm BR}(B_u\rightarrow\tau \nu_{\tau})_{\rm MSSM}}
 {{\rm BR}(B_u\rightarrow \tau \nu_{\tau})_{\rm SM}}
        \leq 2.41 \; (3\sigma)~~&\cite{Asner:2010qj}  ~.~&
\end {eqnarray}
In addition to above constraints we impose the following bounds from the LHC and WMAP9 experiments
\begin{eqnarray}
m_h  = 123-127~{\rm GeV}~~&\cite{ATLAS, CMS} ~,~& \\ 
m_{\tilde{g}} \gtrsim  1.7 \, {\rm TeV}\ ({\rm for}\ m_{\tilde{g}}\sim m_{\tilde{q}}) &
\cite{Chatrchyan:2013wxa, Aad:2014wea}~,~\\  
m_{\tilde{g}}\gtrsim 1.3 \, {\rm TeV}\ ({\rm for}\ m_{\tilde{g}}\ll m_{\tilde{q}}) &
\cite{Chatrchyan:2013wxa, Aad:2014wea}~,~\\
 0.0913 \leq \Omega_{\rm CDM}h^2 (\rm WMAP9) \leq 0.1363  \; (5\sigma)~~&\cite{WMAP9} ~,~&\\
 4.7 \times 10^{-10} \leq \Delta a_{\mu} \leq 52.7 \times 10^{-10} \; (3\sigma)~~&\cite{Bennett:2006fi}~.~&
\end{eqnarray}
\begin{figure}[htp!]
\centering
\subfiguretopcaptrue
\subfigure{
\includegraphics[totalheight=5.5cm,width=7.cm]{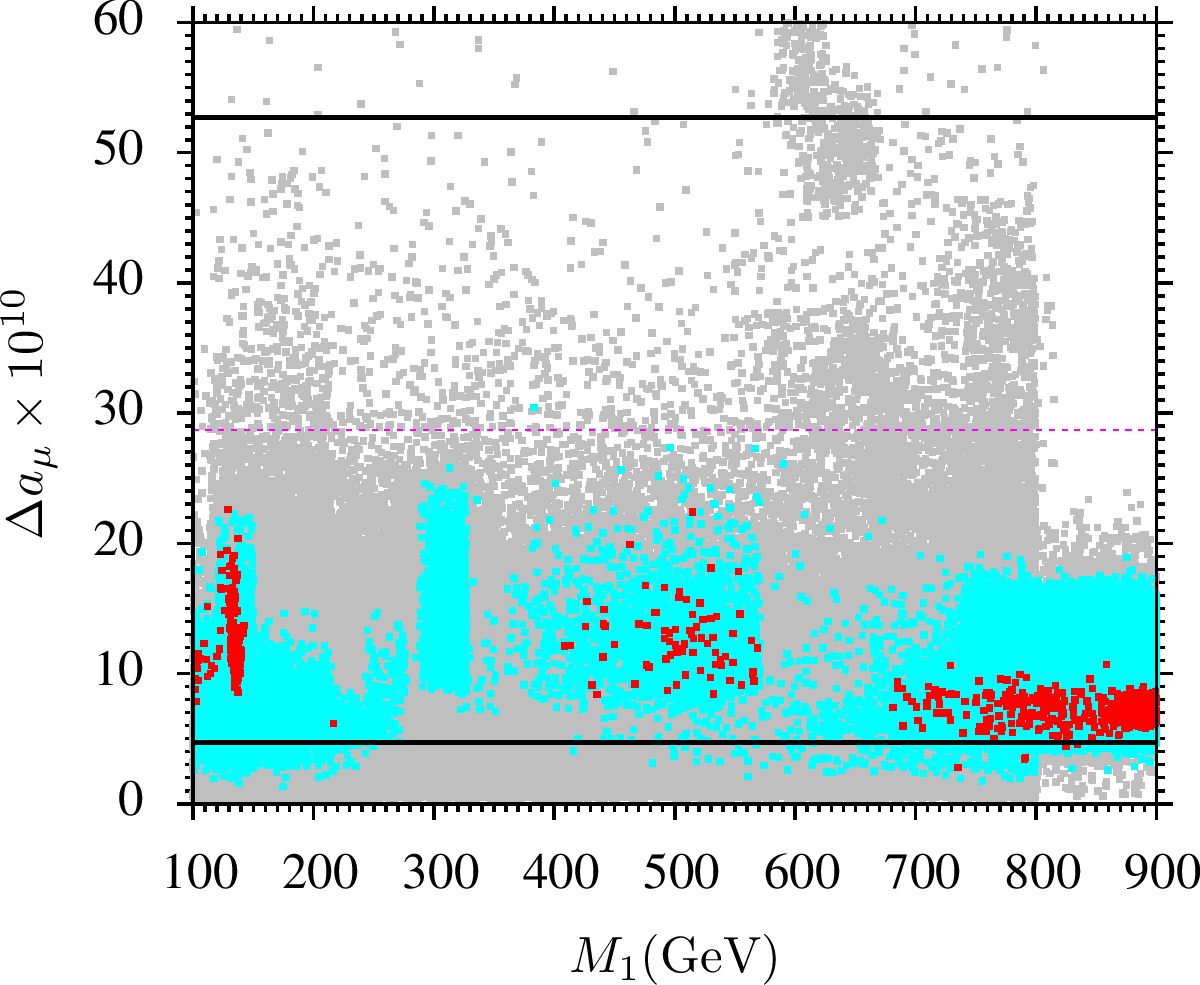}
}
\subfigure{
\includegraphics[totalheight=5.5cm,width=7.cm]{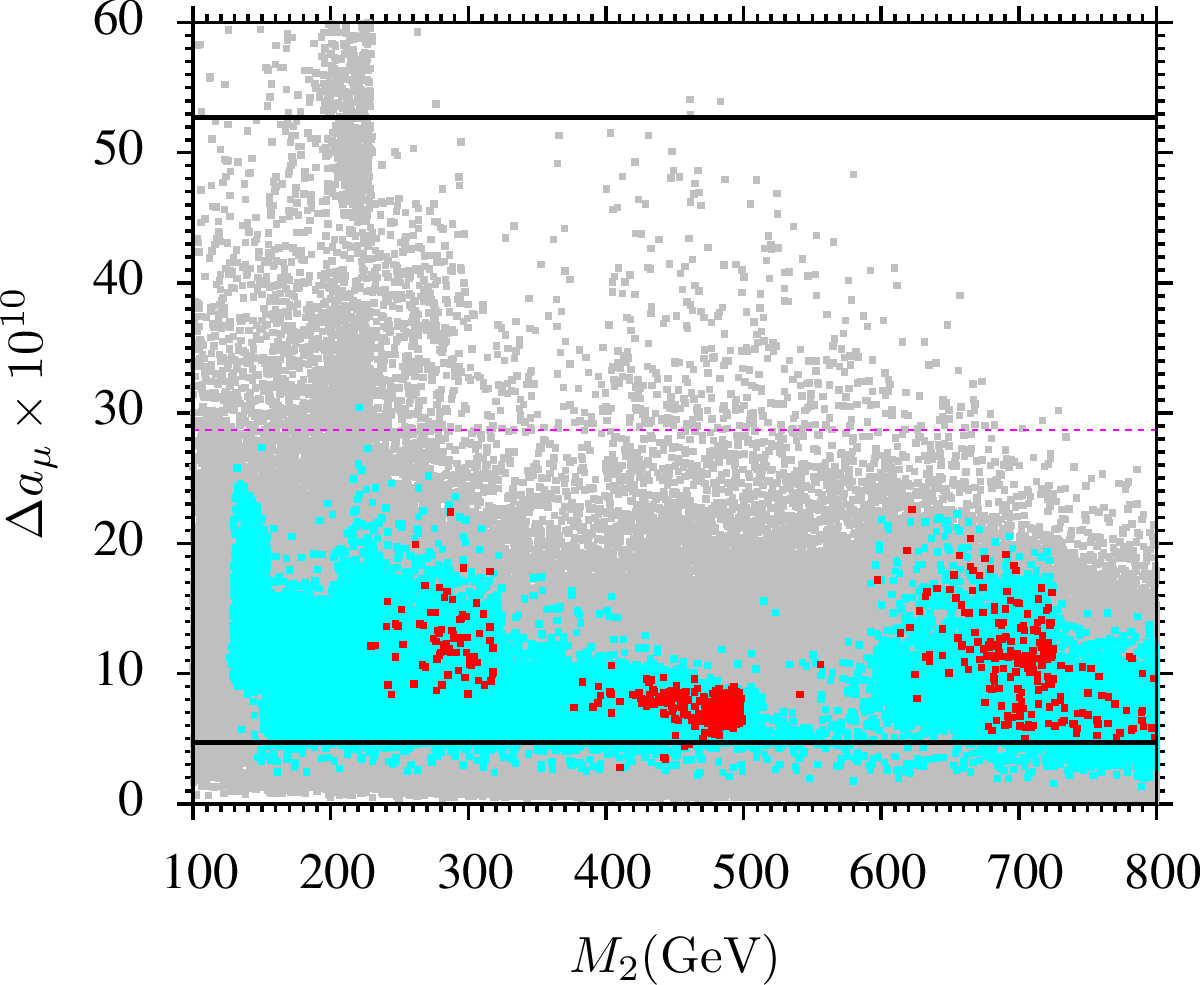}
}
\subfigure{
\includegraphics[totalheight=5.5cm,width=7.cm]{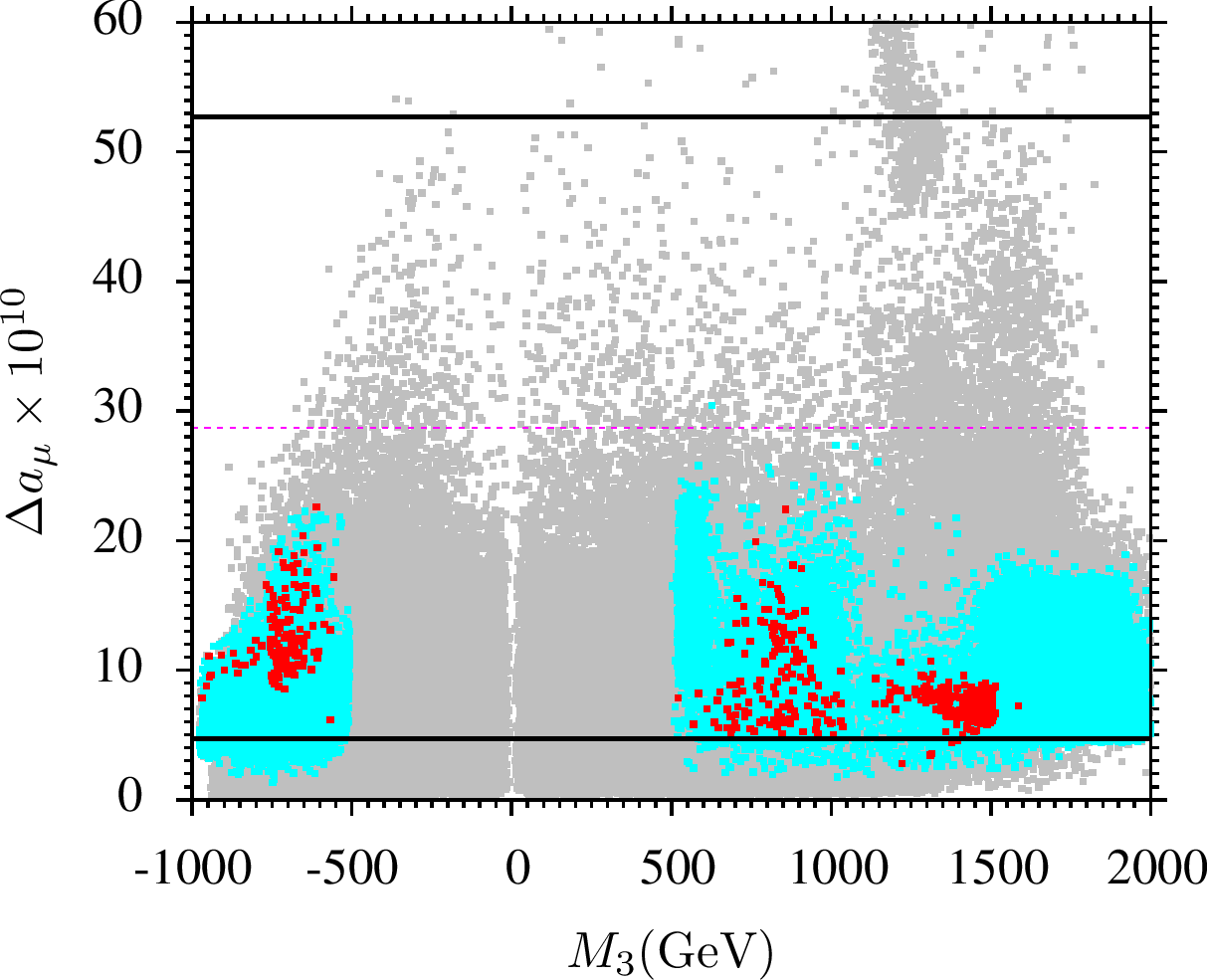}
}
\subfigure{
\includegraphics[totalheight=5.5cm,width=7.cm]{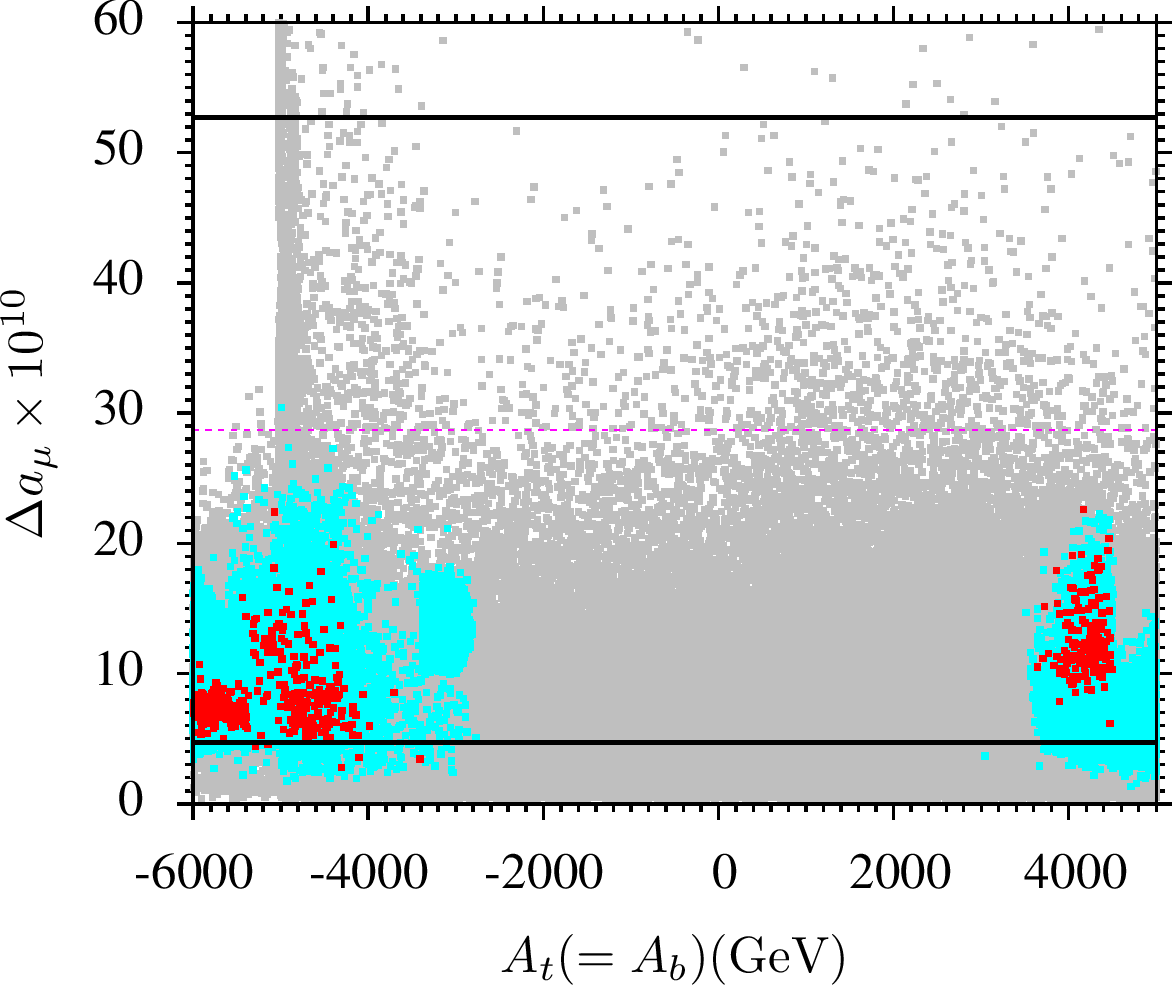}
}
\subfigure{
\includegraphics[totalheight=5.5cm,width=7.cm]{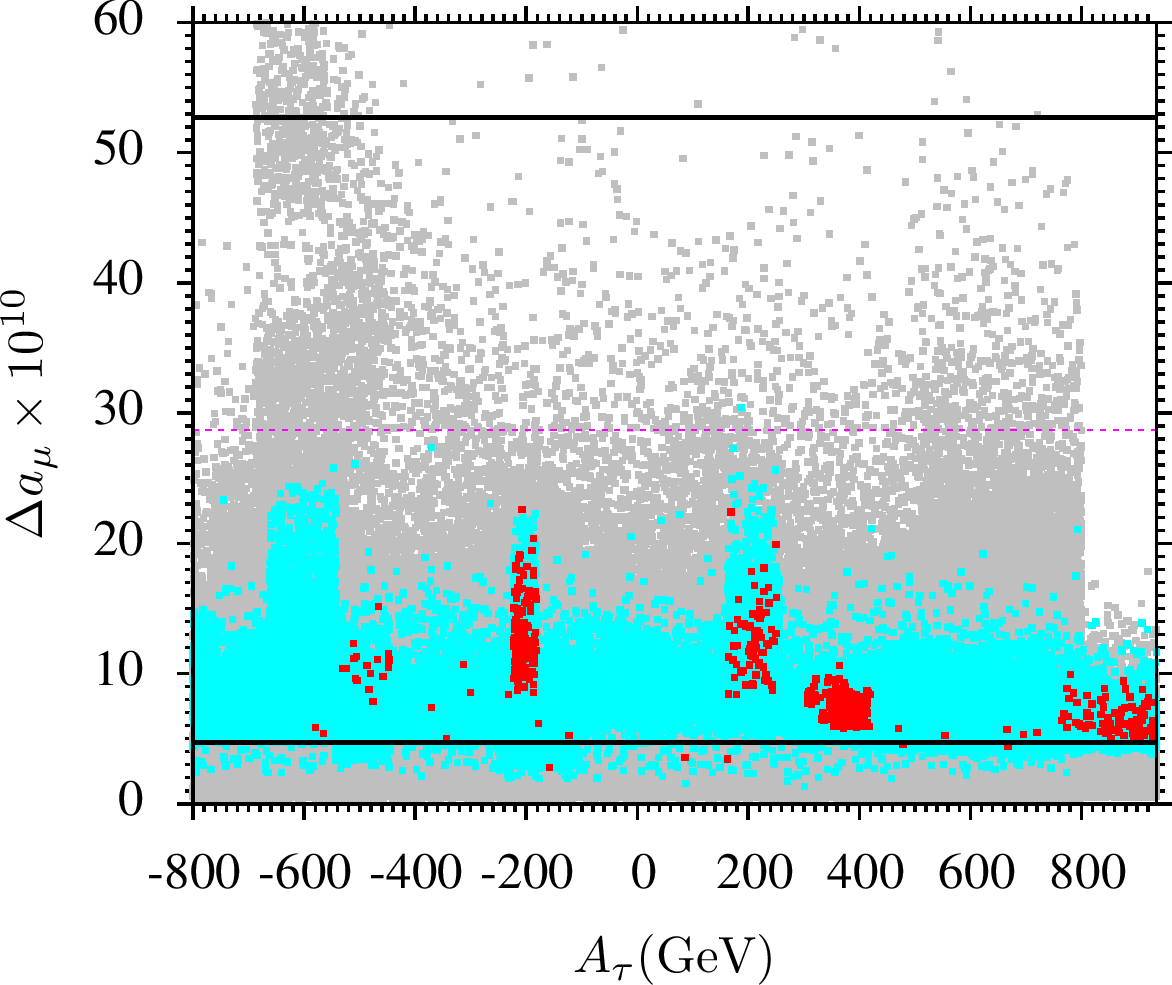}
}
\subfigure{
\includegraphics[totalheight=5.5cm,width=7.cm]{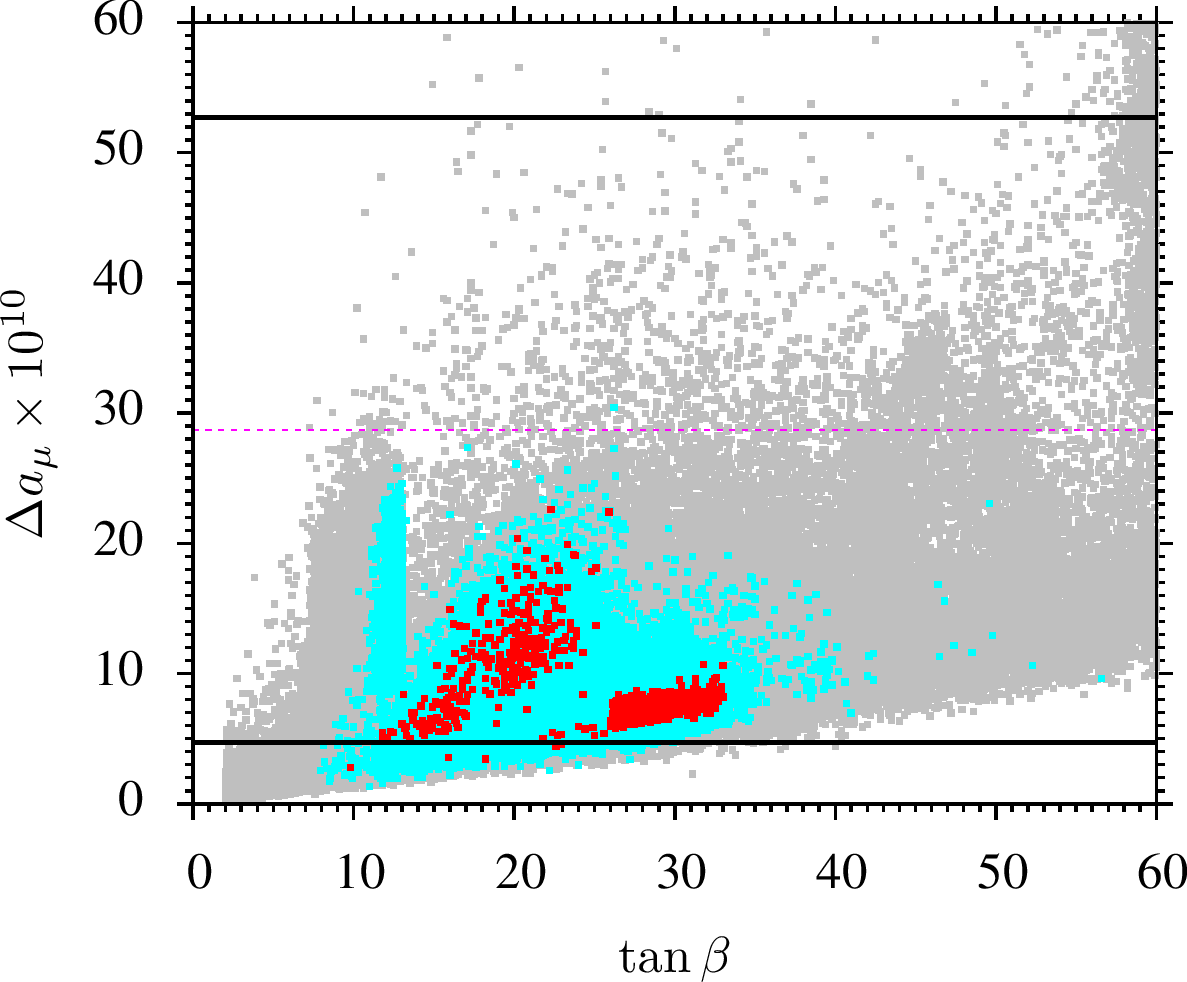}
}

\caption{
Plots in $M_{1}-\Delta a_{\mu}$, $M_{2}-\Delta a_{\mu}$, $M_{3}-\Delta a_{\mu}$, $A_{t}(=A_{b})-\Delta a_{\mu}$, 
$A_{\tau}-\Delta a_{\mu}$ and $\tan\beta-\Delta a_{\mu}$.
Grey points (grey in black and white print) satisfy the REWSB and LSP neutralino  conditions. Aqua points
(slightly grey in black and white print) satisfy the mass bounds, B-physics bounds and $123~{\rm GeV}\leqslant m_h\leqslant 127~{\rm GeV}$.
Red points (dark grey in black and white print) are subset of aqua points that also satisfy the WMAP9 5$\sigma$ bounds.
The horizontal black solid lines represent 3$\sigma$ $\Delta a_{\mu}$ values and 
the dashed purple lines show the central value of $\Delta a_{\mu}$.}
\label{fig1}
\end{figure}
\section{Numerical Results}
\label{results}
\subsection{The Preferred Masses Required by $\Delta a_{\mu}$}
\label {results1}

\begin{figure}[htp!]
\centering
\subfiguretopcaptrue

\subfigure{
\includegraphics[totalheight=5.5cm,width=7.cm]{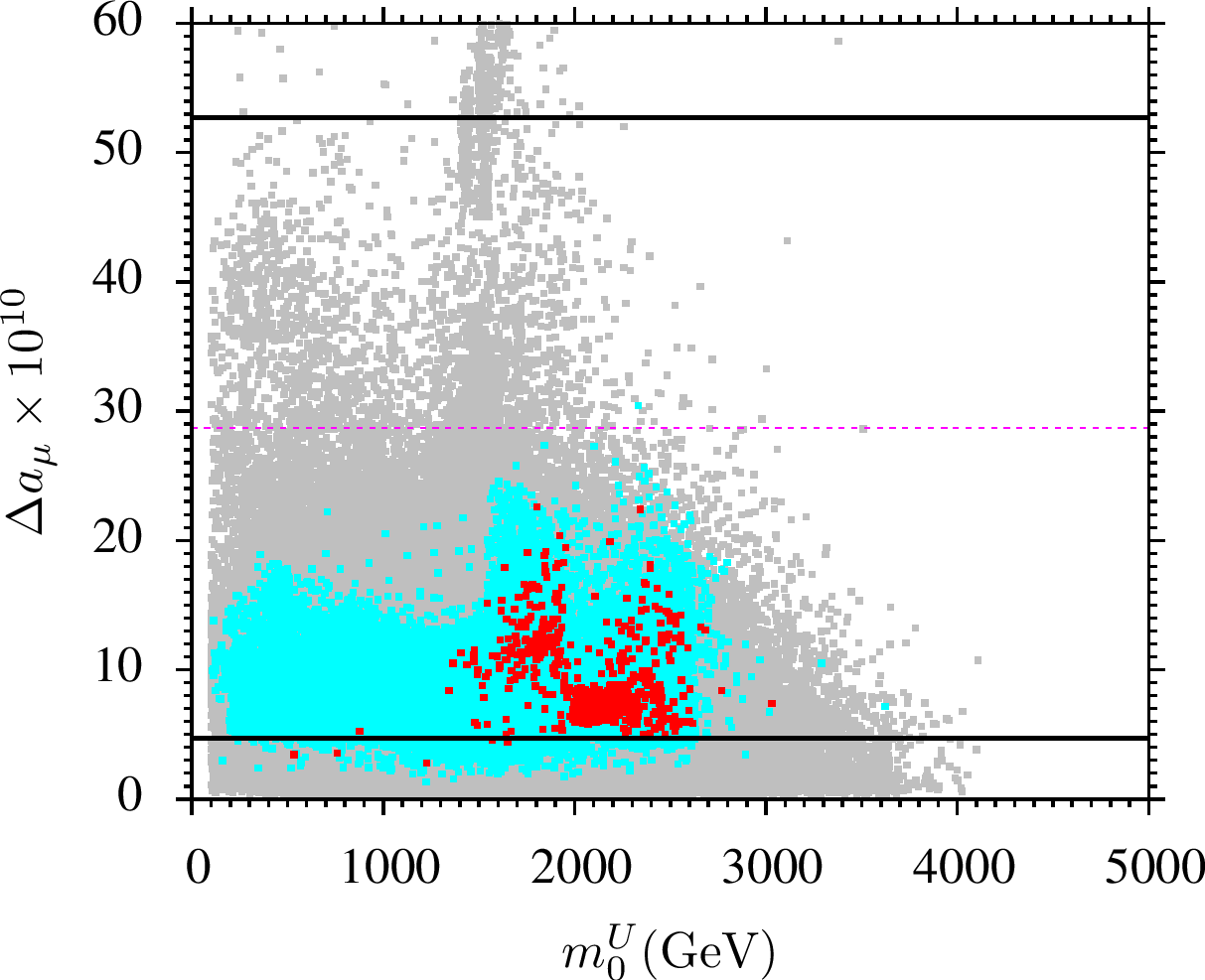}
}
\subfigure{
\includegraphics[totalheight=5.5cm,width=7.cm]{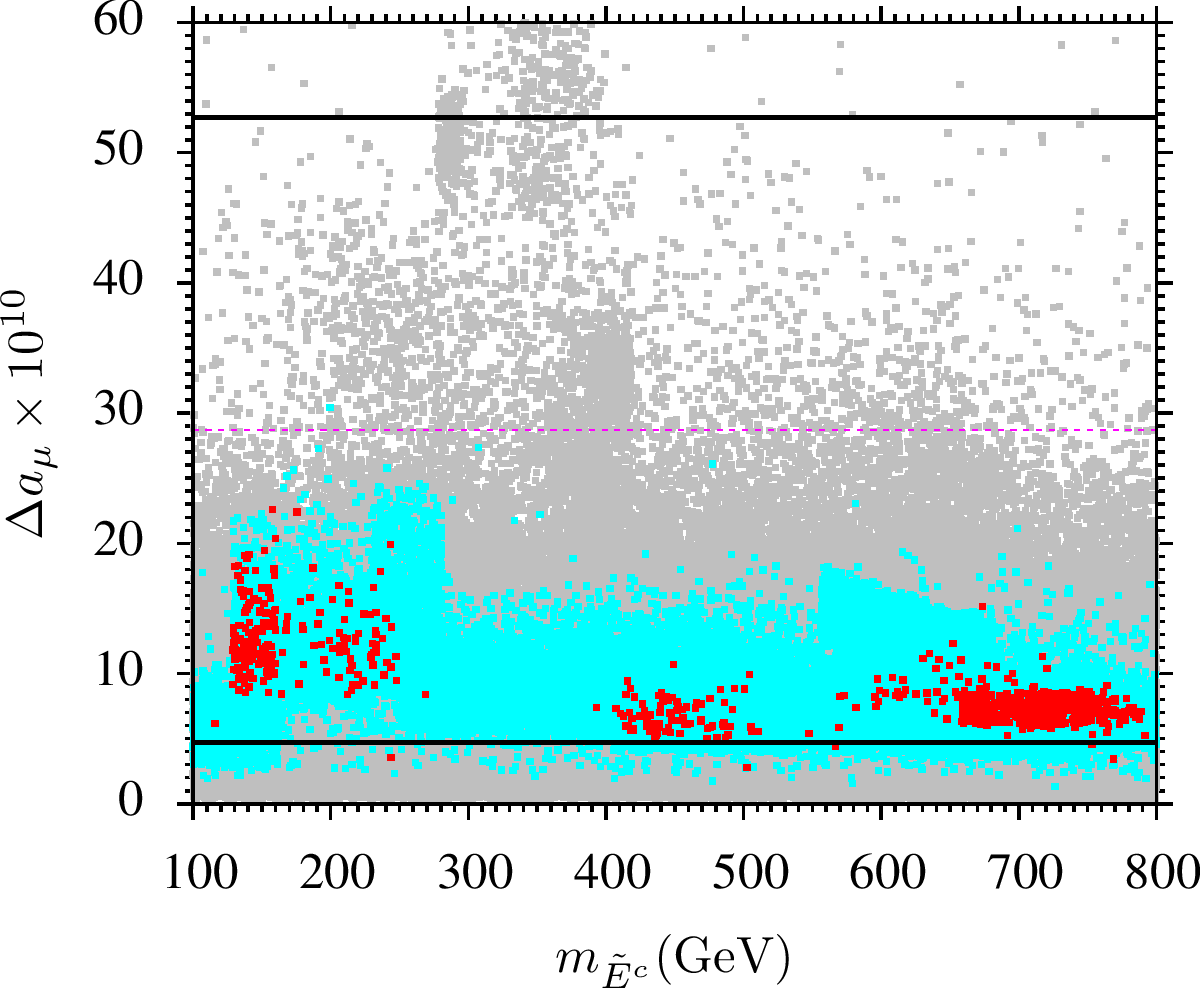}
}
\subfigure{
\includegraphics[totalheight=5.5cm,width=7.cm]{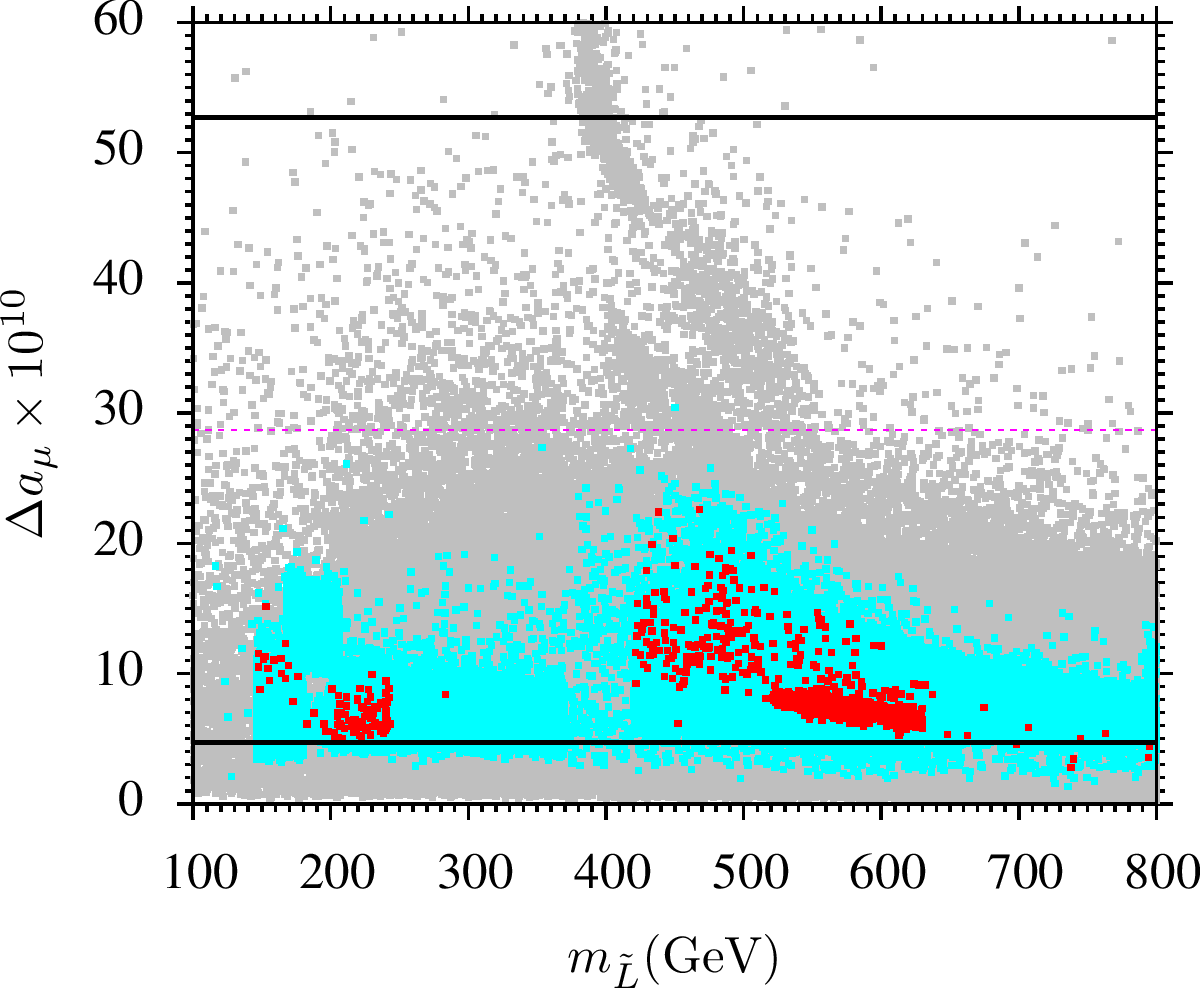}
}
\subfigure{
\includegraphics[totalheight=5.5cm,width=7.cm]{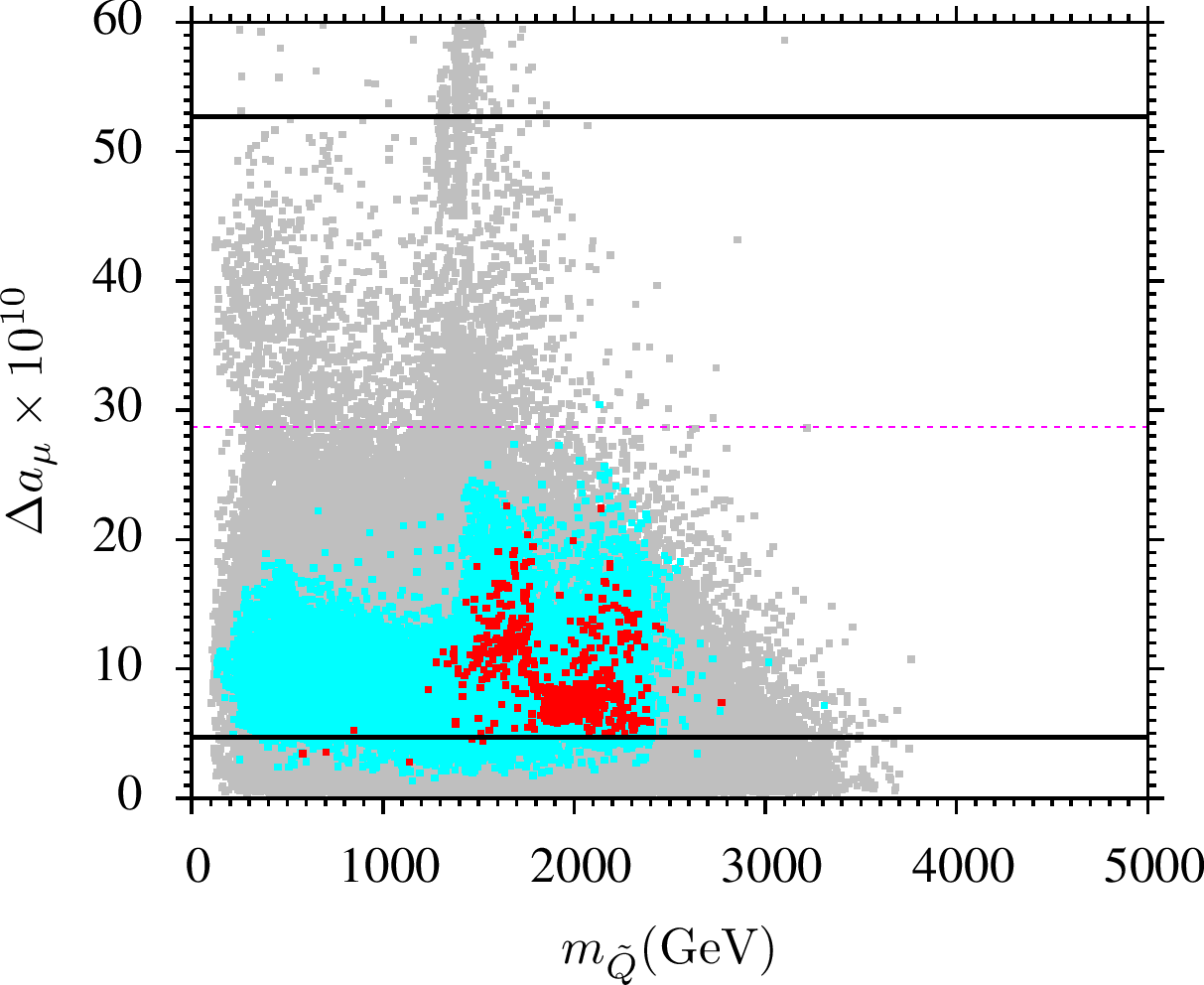}
}
\subfigure{
\includegraphics[totalheight=5.5cm,width=7.cm]{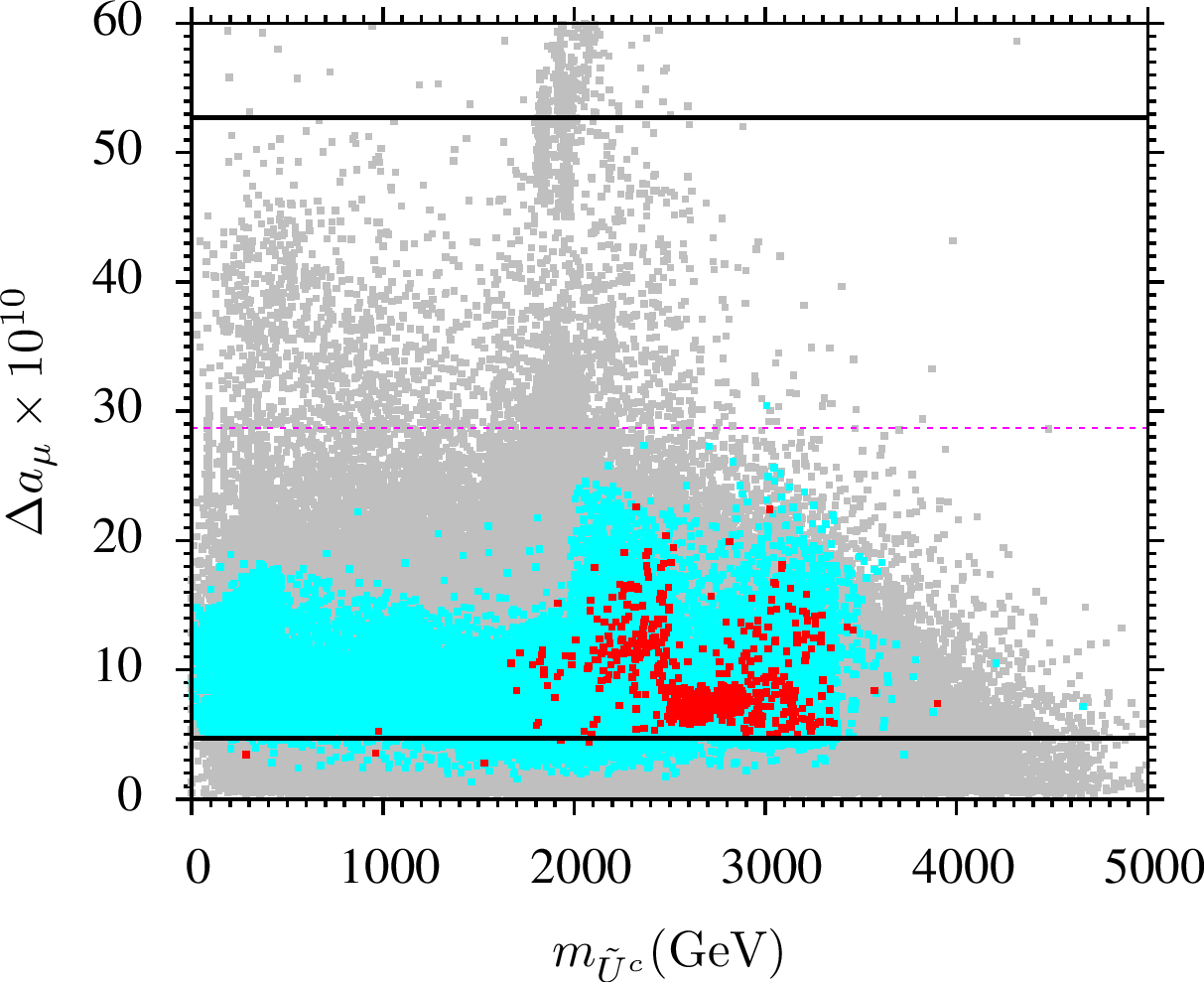}
}
\subfigure{
\includegraphics[totalheight=5.5cm,width=7.cm]{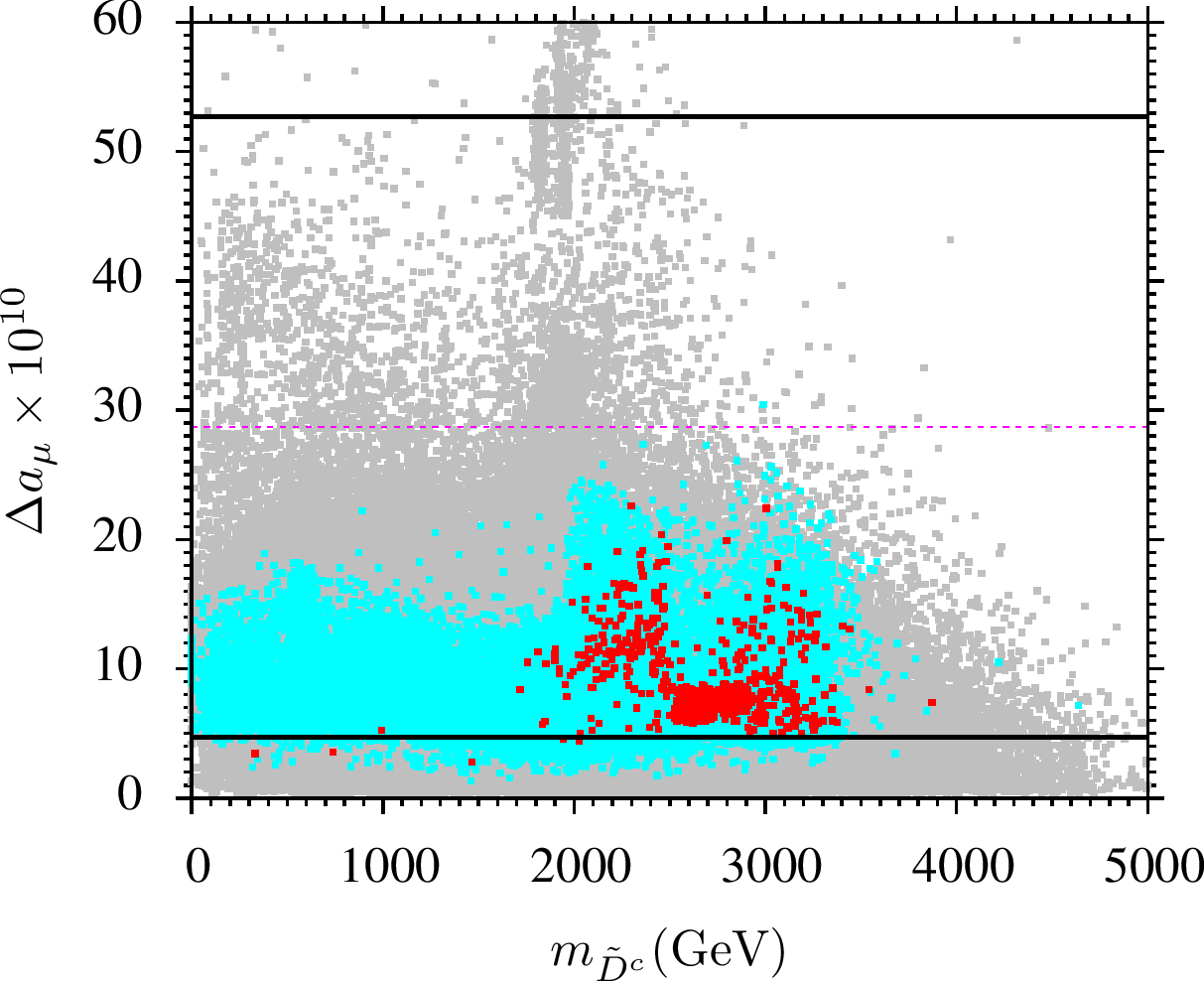}
}

\caption{Plots in $m_{0}^{U}-\Delta a_{\mu}$, $m_{E}-\Delta a_{\mu}$, $m_{L}-\Delta a_{\mu}$, $m_{Q}-\Delta a_{\mu}$, $m_{U}-\Delta a_{\mu}$
and $m_{D}-\Delta a_{\mu}$ planes.
The color coding is the same as in Fig.~\ref{fig1}.
}
\label{fig2}
\end{figure}

In this subsection we present results of our scans. 
In Figs.~\ref{fig1}-\ref{fig2} we present graphs of $\Delta a_{\mu}$ 
versus the input parameters given in Section~\ref{sec:scan}.
In these plots, grey points (grey in black and white print) satisfy the REWSB and LSP neutralino  conditions, 
aqua points (slightly dark grey in black and white print) satisfy the mass bounds, B-physics bounds, 
and $123~{\rm GeV}\leqslant m_h\leqslant 127~{\rm GeV}$,
and red points (dark grey in black and white print) are subset of aqua points that also satisfy the WMAP9 5$\sigma$ bounds.
In Fig.~\ref{fig1}, we display graphs in $M_{1}-\Delta a_{\mu}$, 
$M_{2}-\Delta a_{\mu}$, $M_{3}-\Delta a_{\mu}$, $A_{t}(=A_{b})-\Delta a_{\mu}$, $A_{\tau}-\Delta a_{\mu}$, 
and $\tan\beta-\Delta a_{\mu}$ planes. At first we did general scans over the parameter space given by 
Eq.~(\ref{input_param_range}) and then we did the dedicate
scans around the phenomenologically interesting solutions. These dedicated searches appear 
as patches in the graphs. In the 
top left panel we see that aqua points have $M_1$ mass range $[100,~900]$ GeV 
which also have 3$\sigma$ to 1$\sigma$
contributions to $\Delta a_{\mu}$. There is a lack of grey points between 
$800\, \rm GeV \lesssim M_1 \lesssim 900 \, \rm GeV$. It is because initially 
we generated data up to $M_{1}=800$ GeV.  
In order to get the light CP-even Higgs boson mass around 125 GeV, we then
 did some dedicated searches where we had to increase upper ranges of a couple of input parameters.
This is the reason why one can see the sharp cut in grey points 
in this plot and plots in $A_{t}(=A_{b})-\Delta a_{\mu}$ and $A_{\tau}-\Delta a_{\mu}$ planes.
Another point to be noted is that we do not see any preferred range of $M_1$ 
to have large contribution to $\Delta a_{\mu}$.
Apparently, there are more points between $400\, \rm GeV \lesssim M_1 \lesssim 800 \, \rm GeV$ 
where we see large values for $\Delta a_{\mu}$. But in fact by generating more data 
it can be shown that 
we have more or less same
contributions to $\Delta a_{\mu}$ for all values of $M_1$ between $[100,900]$ GeV. 
Since our parameter space is very large
so doing these kind of scans is a very time consuming job. But the main purpose of this study 
 is to show that the EWSUSY from
GmSUGRA can resolve the apparent discrepancy of muon $\Delta a_{\mu}$, which is clearly displayed. 
Moreover, we find that even red points can have any value of $M_1$ between $[100,~900]$ GeV. We note that red points with small $M_1$ values 
($M_{1}\lesssim$ 150 GeV) with 2$\sigma$ contributions to $\Delta a_{\mu}$ represent resonance solutions like $Z$-pole and Higgs-pole. 
We will discuss such solution in Section~\ref{dm}. We also
see that in our present data with $M_1 \sim$ 500 GeV, red points have contributions 
to $\Delta a_{\mu}$ within 1$\sigma$. In the top right panel, we note that aqua points 
can have $M_2$ values between 140-800 GeV and within 3$\sigma$ 
bounds of $\Delta a_{\mu}$. The small values of $M_2$ indicate the presence of light wino-type LSP 
neutralino. On the other hand, the minimal and maximal
 $M_2$ values for red points are between 250 GeV and 800 GeV. 
In the middle left panel we display values for $M_3$ which we 
calculate using Eq.~(\ref{M3}). Here one can see that we have solutions with
both $M_3 <0$ and $M_3>0$. In order to have 3$\sigma$ or better $\Delta a_{\mu}$ contributions and remain consistent with the constraints discussed in Section~\ref{sec:scan}, we need in both cases $|M_3| >$  500 GeV and 
which indicates 
relatively heavy gluino. The right middle panel depicts that in our model 
with $|A_t|=|A_b| > $ 3000 GeV, we have 
the sizable SUSY contribution to $\Delta a_{\mu}$ and consistent with the bounds given in Section~\ref{sec:scan}. 
These relatively large values of $|A_t|$  will  also help to get
Higgs boson mass around 125 GeV. In the left bottom panel 
we observe that aqua solutions have $A_{\tau}$ range anywhere
between $-800 \, {\rm GeV} \lesssim A_{\tau} \lesssim 935 \, {\rm GeV}$. But for red points we have
$-600 \, {\rm GeV} \lesssim A_{\tau} \lesssim 935 \, {\rm GeV}$. In the bottom right panel
we see that the contributions to $\Delta a_{\mu}$ increase as $\tan\beta$ increases, which can be understood 
from Eq.~(\ref{amu_susy}). For $\tan\beta \approx$ 12-50 and 20-25 respectively for
 aqua and red points, we have 
solutions within 1$\sigma$ ($20.7\times 10^{-10}-36.7\times 10^{-10}$) bounds on $\Delta a_{\mu}$. 
As we discussed earlier, the large $\tan\beta$ along
with large $\mu$ values may help getting the desired $\Delta a_{\mu}$ values. 
But the large left-right stau mixing term $A_{\tau}$ may generate the electric 
charge breaking minimum in the scalar potential as indicated in \cite{Rattazzi:1996fb}. 
It was shown in \cite{Hisano:2010re} that one can have a 
metastability condition for the electric charge breaking in terms of $\mu$, $\tan\beta$, $m_{\tilde \tau_{L}}$ and $m_{\tilde \tau_{R}}$ (also see \cite{Sato:2012bf}), where the product $\mu\tan\beta$ should be less than some combination of $m_{\tilde \tau_{L}}$ and $m_{\tilde \tau_{R}}$. 
Although in our case we do not have the very large $A_{\tau}$ values, we still use Eq.~(11) of \cite{Hisano:2010re} to 
filter out points which do not satisfy the metastability condition.

In Fig.~\ref{fig2} we show plots in $m_{0}^{U}-\Delta a_{\mu}$, 
$m_{\tilde E^c}-\Delta a_{\mu}$, $m_{\tilde L}-\Delta a_{\mu}$, 
$m_{\tilde Q}-\Delta a_{\mu}$, $m_{\tilde U^c}-\Delta a_{\mu}$, and $m_{\tilde D^c}-\Delta a_{\mu}$ planes. 
The color coding is the same as in Fig.~\ref{fig1}.   
In top left panel we see that $m_{0}^{U}$ is any where between 100-3600 GeV 
if we consider aqua points but for red point it is 
restricted to be around 3000 GeV. In the right top panel 
we observe that aqua points within 1-3$\sigma$ bounds on 
$\Delta a_{\mu}$ have $m_{E^c}$ from 100 to 800 GeV. Similarly, red points share the same mass range. 
The middle left
plot shows the mass range $[130,~800]$~GeV for the universal left-handed sleptons $\tilde L$. Like 
the right-handed sleptons $\tilde E^c$, the left-handed sleptons $\tilde L$
more or less share the same mass range for both aqua and red points. In the right middle, bottom left and 
bottom right panels we display masses for left-handed, right-handed up-type and down-type squarks, respectively,
 which we calculate by using Eq.~(\ref{squarks_masses}). For the left-handed squarks we have slightly narrow
allowed mass ranges as compared 
to the right-handed squarks.
Also, $\tilde U^c$ and $\tilde D^c$ have almost the same mass ranges, which are consistent with 
Eq.~(\ref{squarks_masses}). 

\subsection{Compatibility Between  the $\Delta a_{\mu}$ Bound and EWFT}
\label {results2}

\begin{figure}[htp!]
\centering
\subfiguretopcaptrue

\subfigure{
\includegraphics[totalheight=5.5cm,width=7.cm]{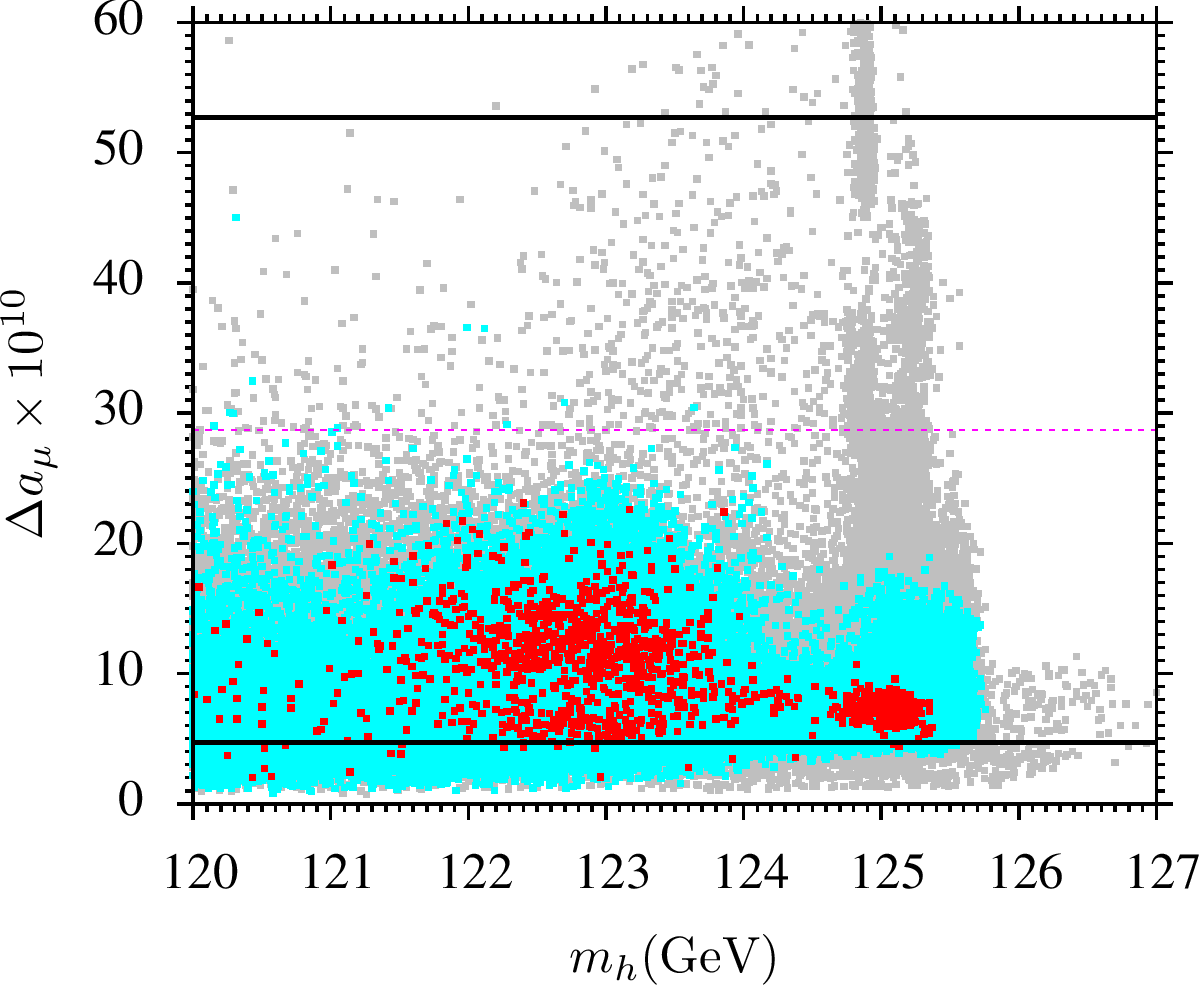}
}
\subfigure{
\includegraphics[totalheight=5.5cm,width=7.cm]{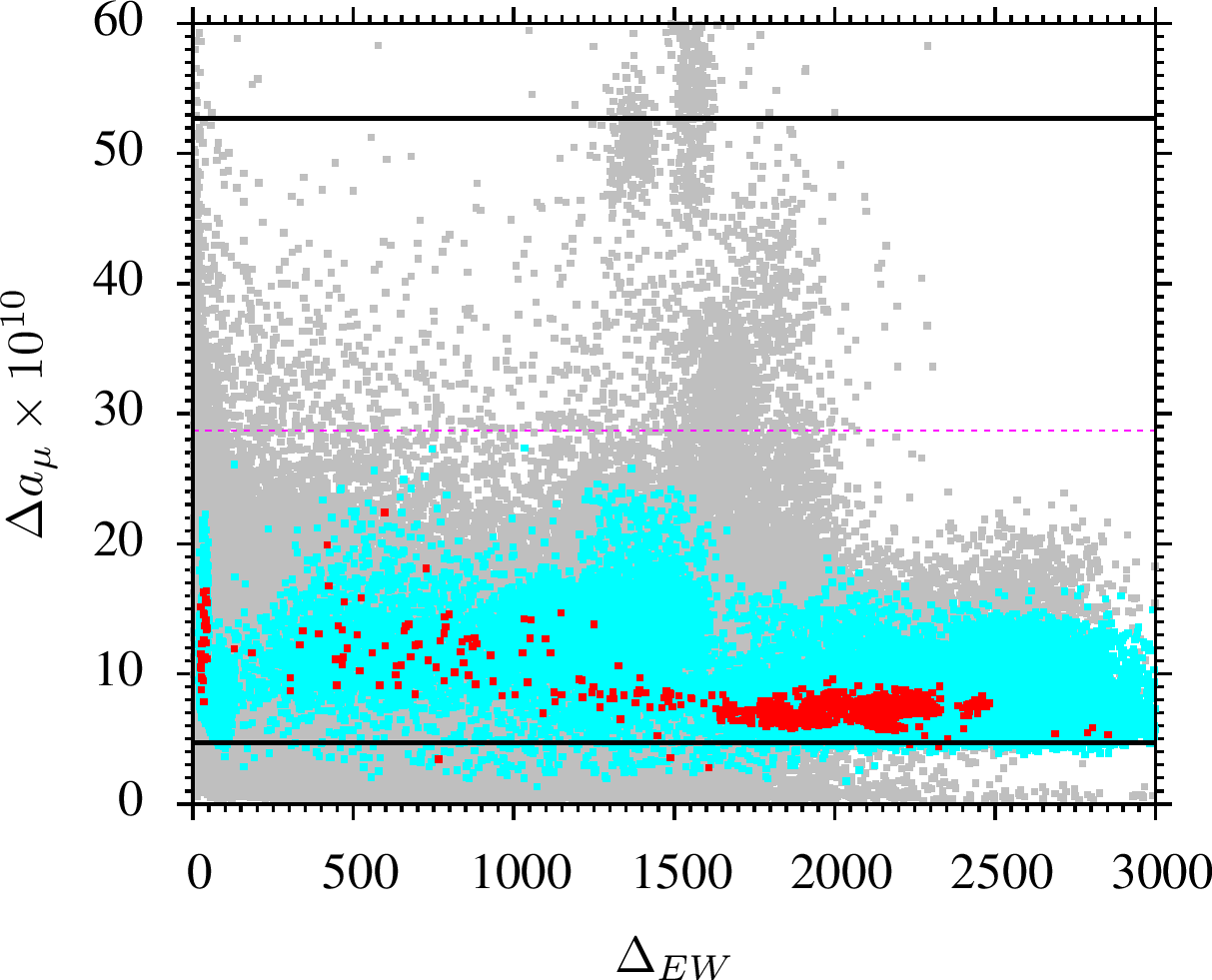}
}
\subfigure{
\includegraphics[totalheight=5.5cm,width=7.cm]{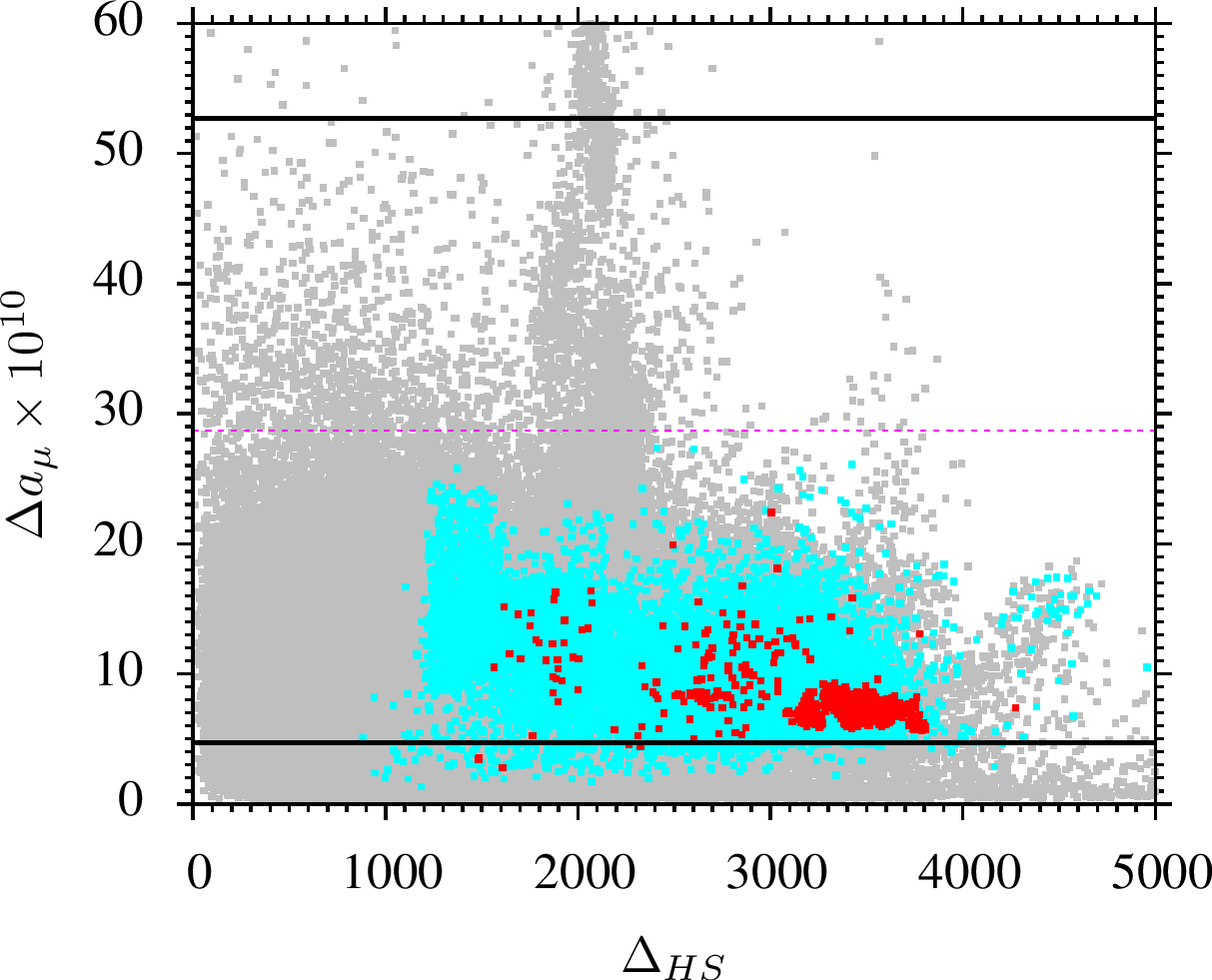}
}
\caption{Plots in $m_{h}-\Delta a_{\mu}$, $\Delta_{EW}-\Delta a_{\mu}$ and $\Delta_{HS}-\Delta a_{\mu}$ planes.
The color coding is the same as in Fig.~\ref{fig1} except that 
in $m_{h}-\Delta a_{\mu}$ plane we do not apply the Higgs boson mass
bound.
}
\label{fig3}
\end{figure}

Fig.~\ref{fig3} displays plots in $m_{h}-\Delta a_{\mu}$, $\Delta_{EW}-\Delta a_{\mu}$ and $\Delta_{EW}-\Delta a_{\mu}$ planes.
The color coding is the same as in Fig.~\ref{fig1} except that 
in $m_{h}-\Delta a_{\mu}$ plane we do not apply Higgs mass 
bound. The left panel shows plenty of
solutions accommodating bounds on Higgs boson mass 123-127 GeV, having
sizable contributions to $\Delta a_{\mu}$, and being consistent with the
 sparticle mass bounds and B-physics bounds mentioned in 
Section~\ref{sec:scan}. Red points are mostly concentrated 
in the Higgs boson mass range  $123~{\rm GeV}\lesssim m_h \lesssim 124~{\rm GeV}$ and have $\Delta a_{\mu}$ contributions
within 2$\sigma$. In the
left panel, we find that aqua and red points have small values for fine-tuning
measure $\Delta_{EW}$. These points not only resolve $\Delta a_{\mu}$ discrepancy but also provide solution
to the EWFT problem. As we mentioned earlier, this can be understood from Eq.~(\ref{amu_susy}), where we see that the SUSY 
contribution to $\Delta a_{\mu}$ is proportional to $\mu$, gaugino masses ($M_{1,2}$) and $\tan\beta$, but inversely 
proportional to the fourth power of $m_{SUSY}$ (the mass scale related to charginos, smuons, sneutrino, and neutralinos). 
Small EWFT requires small values of $\mu$, while sizable
$\Delta a_{\mu}$ contributions have opposite requirement for $\mu$. But if the gaugino masses and $\tan\beta$ are 
appropriately large (as can be seen in Figs.~\ref{fig1}) and $m_{SUSY}$ is small 
(as can be seen below in Fig.~\ref{fig4}), one can indeed have
sizable $\Delta a_{\mu}$. In our data the minimal value of $\Delta_{EW}$ is about 16.5 (6$\%$) with $\Delta a_{\mu}\approx 11.6\times 10^{-10}$ for aqua points and for red points $\Delta_{EW}$ can be as small as 25 (4$\%$) with $\Delta a_{\mu}\approx 15.1\times 10^{-10}$. A plot in $\Delta_{HS}-\Delta a_{\mu}$ is also shown for the comparison of $g-2$ with
the HS fine-tuning measure. Here, it can be seen that
the points, which have low $\Delta_{HS}$ values, have relatively large $\Delta a_{\mu}$ values, and vice versa.

\begin{figure}[htp!]
\centering
\subfiguretopcaptrue
\subfigure{
\includegraphics[totalheight=5.5cm,width=7.cm]{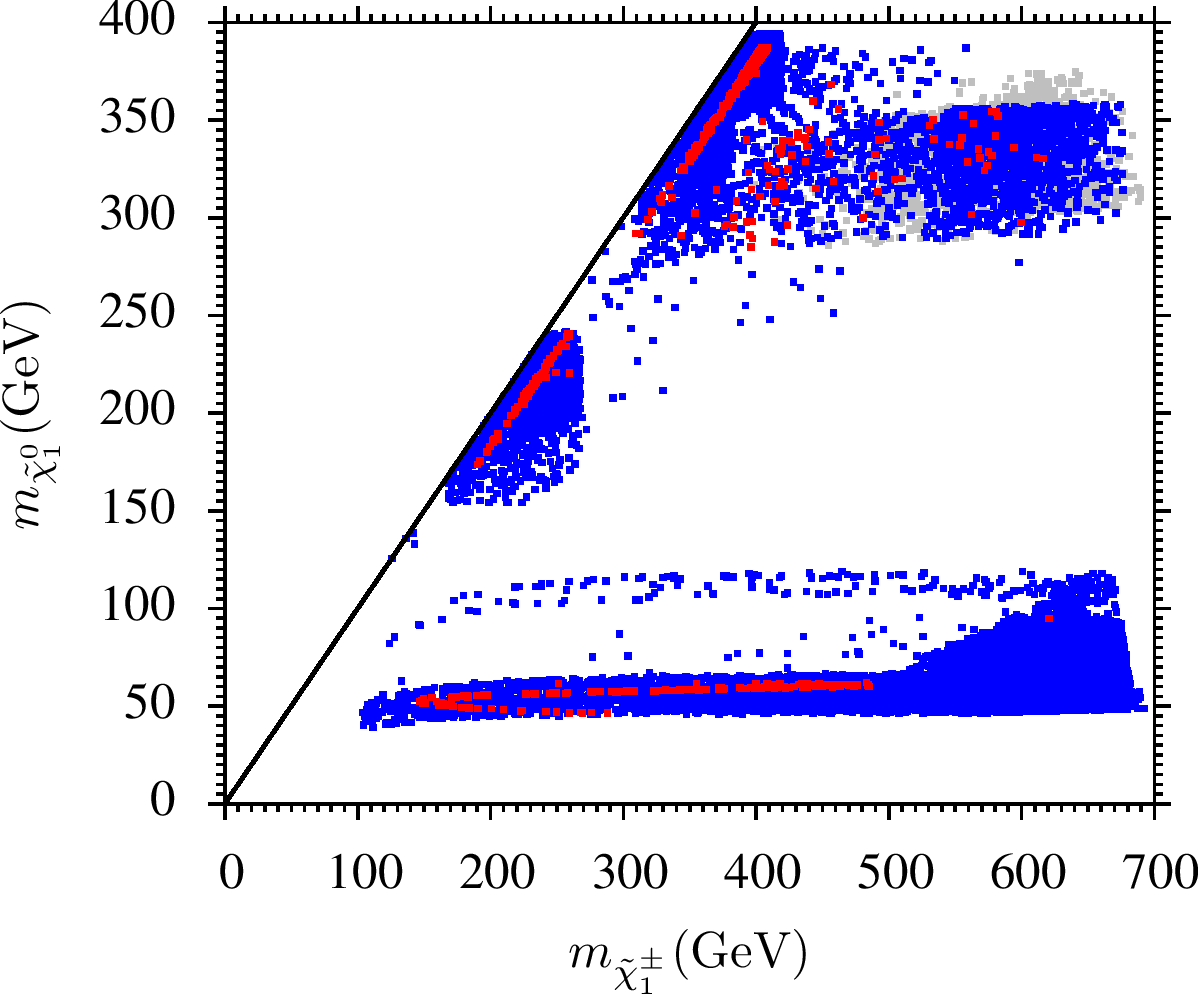}
}
\subfigure{
\includegraphics[totalheight=5.5cm,width=7.cm]{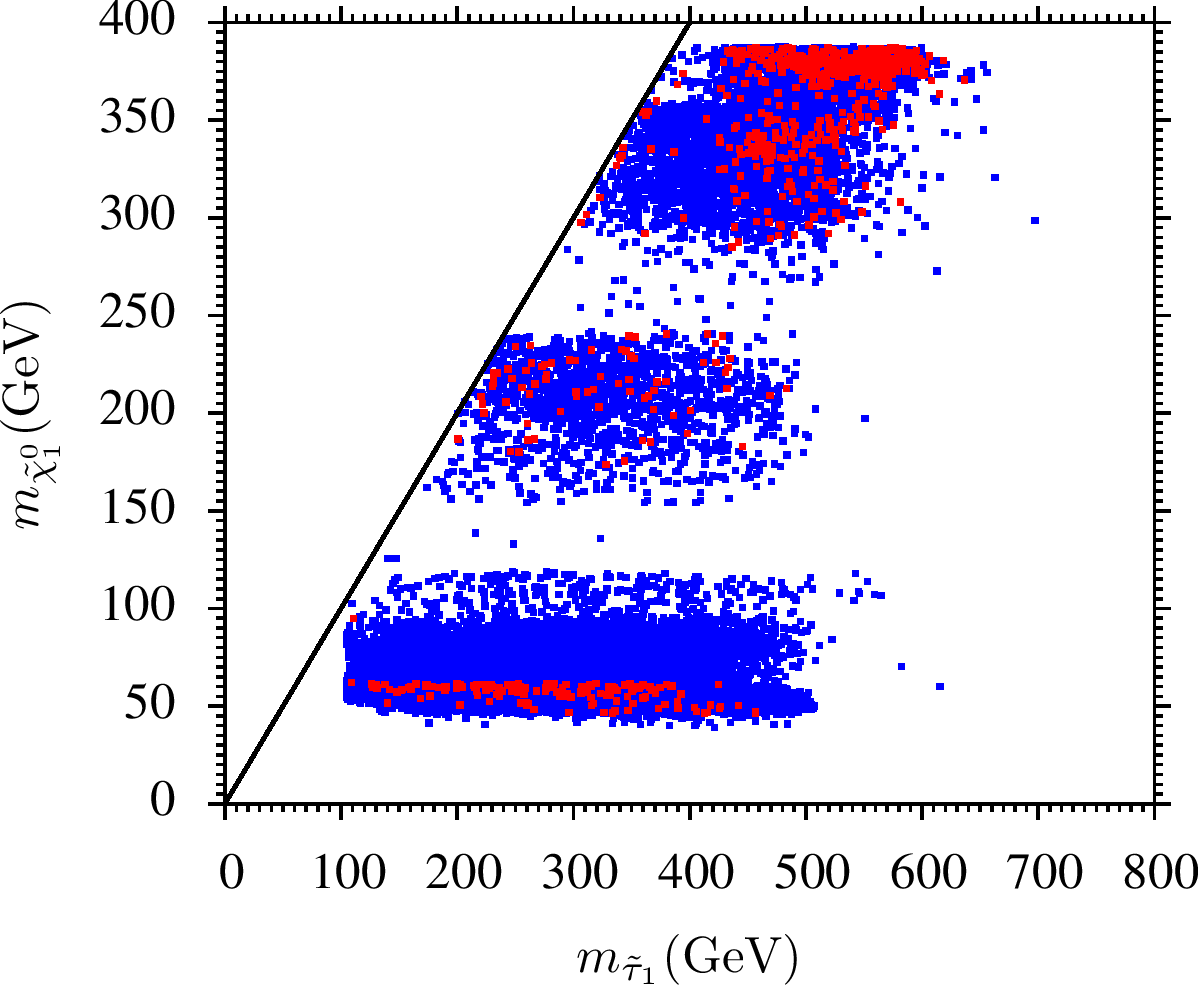}
}
\subfigure{
\includegraphics[totalheight=5.5cm,width=7.cm]{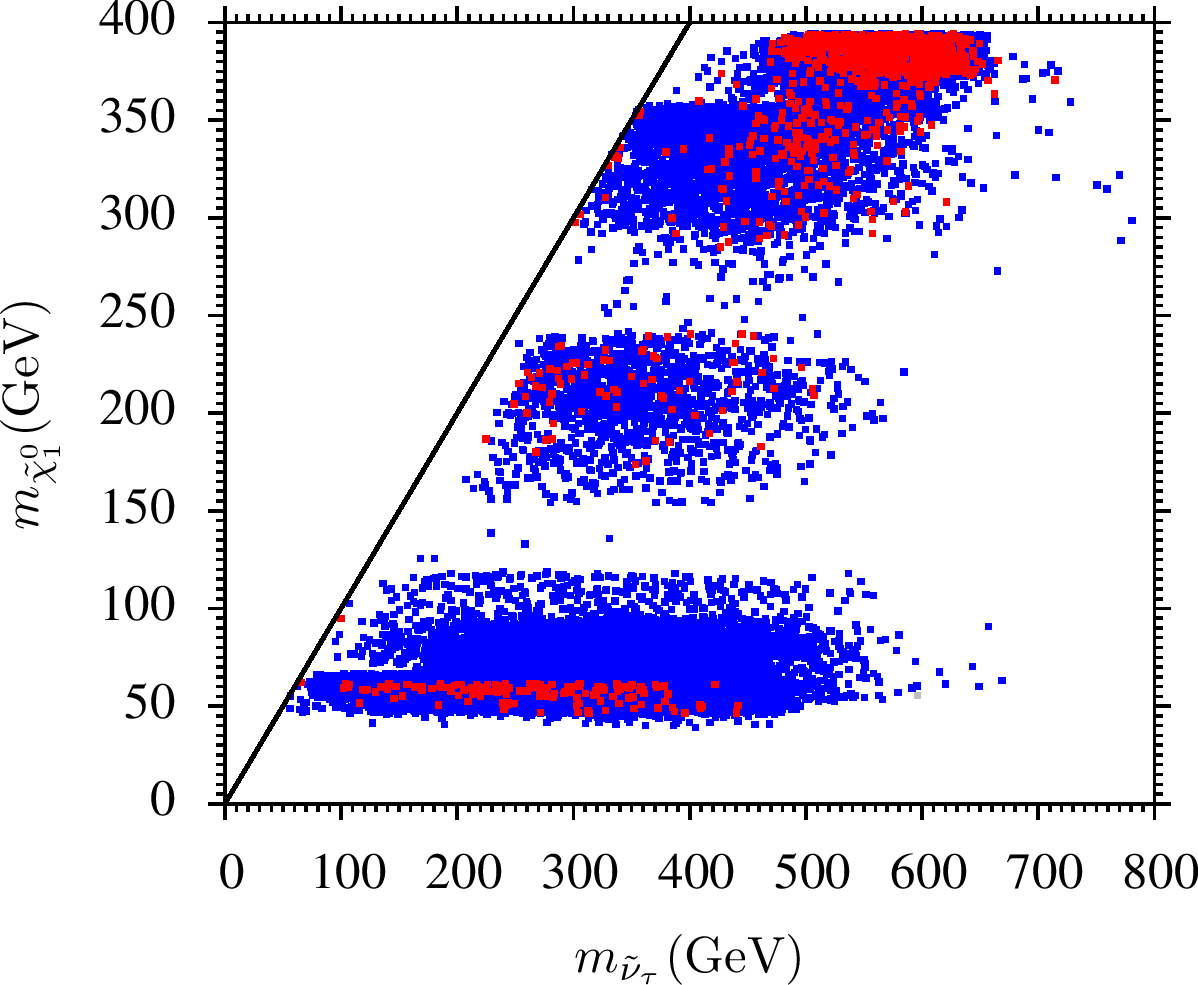}
}
\subfigure{
\includegraphics[totalheight=5.5cm,width=7.cm]{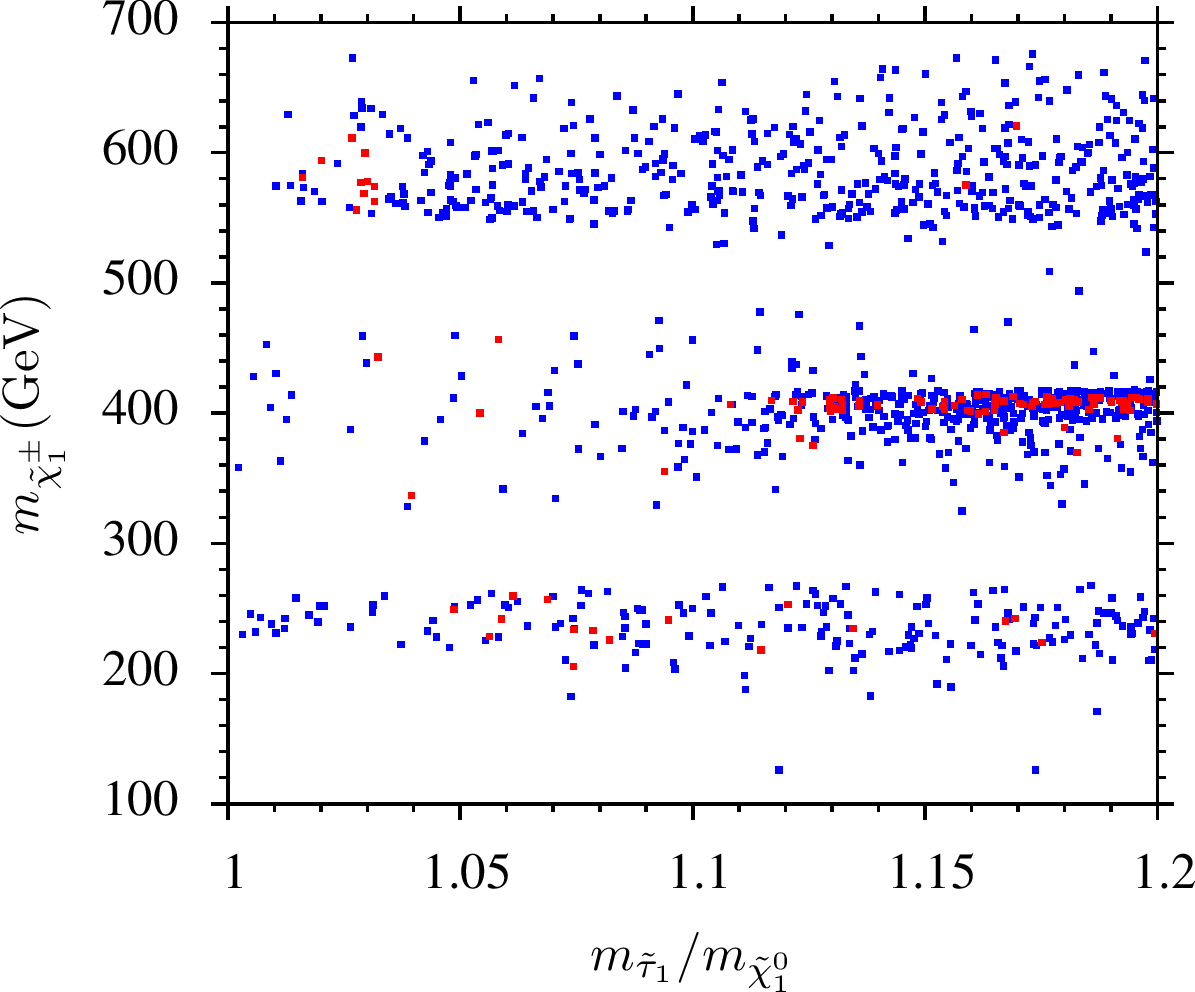}
}
\subfigure{
\includegraphics[totalheight=5.5cm,width=7.cm]{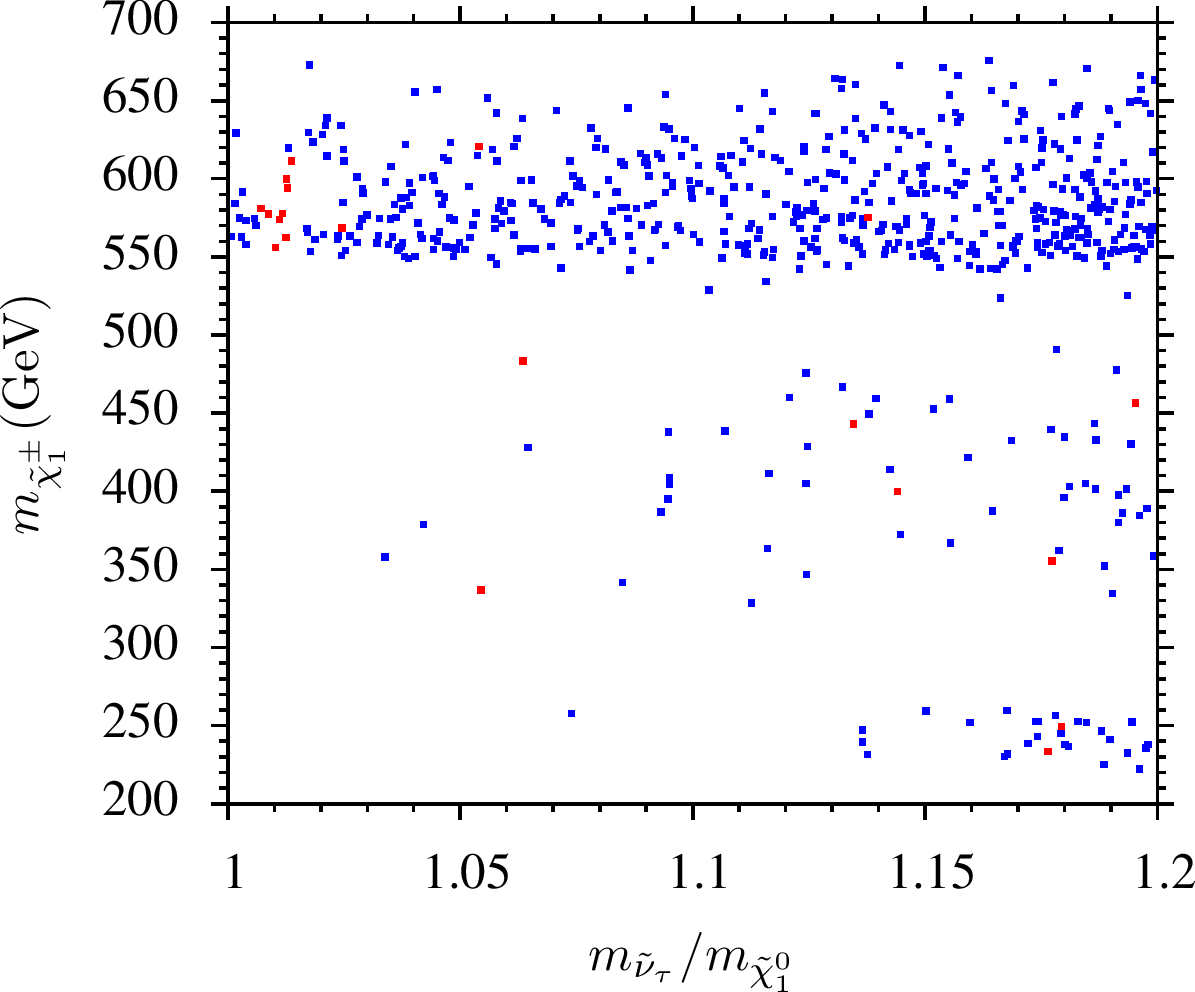}
}
\subfigure{
\includegraphics[totalheight=5.5cm,width=7.cm]{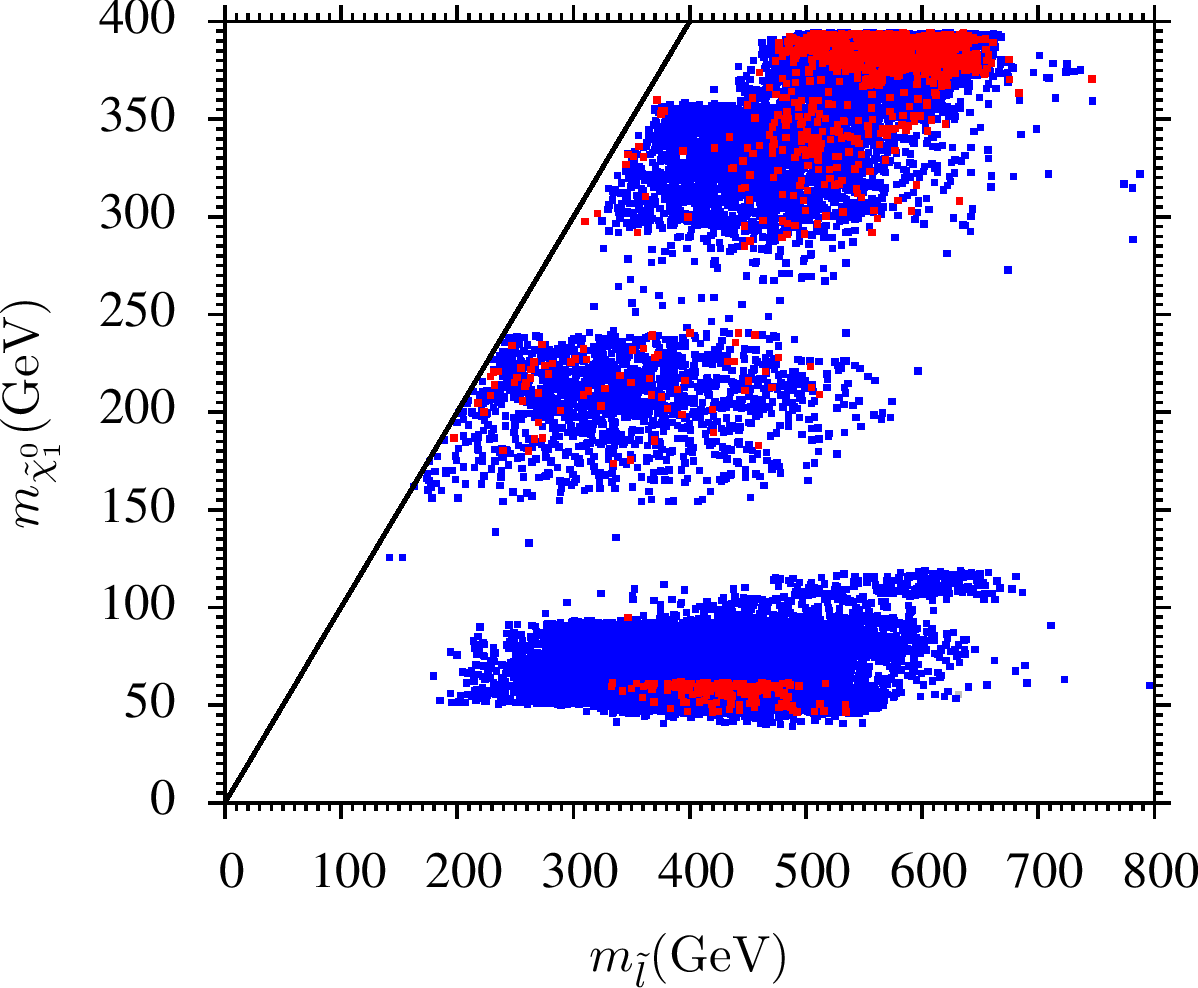}
}
\caption{Plots in $m_{\tilde \chi_{1}^{\pm}}-m_{\tilde \chi_{1}^{0}}$, $m_{\tilde \tau_1}-m_{\tilde \chi_{1}^{0}}$, $m_{\tilde \nu_{\tau}}-m_{\tilde \chi_{1}^{0}}$, $m_{\tilde \tau_1}/m_{\tilde \chi_{1}^{0}}-m_{\tilde \chi_{1}^{\pm}}$, $m_{\tilde \nu_{\tau}}/m_{\tilde \chi_{1}^{0}}-m_{\tilde \chi_{1}^{\pm}}$, and $m_{\tilde l}-m_{\tilde \chi_{1}^{0}}$ planes.
Blue points (black points in black and white print) represent the bino-type LSP neutralino and satisfy REWSB. 
They are consistent with bounds on sparticle and Higgs boson masses
including $123~{\rm GeV}\leqslant m_h\leqslant 127~{\rm GeV}$, B-physics,
 and $4.7\times 10^{-10} \lesssim \Delta a_{\mu} \lesssim 52.7\times 10^{-10}$. 
Red points (dark grey points in black and white print) are subset of blue points which satisfy the WMAP9 5$\sigma$ bound.
}
\label{fig4}
\end{figure}
\subsection{Supersymmetry Searches at the LHC}
\label{results3}
 
The viable parameter space in the SSMs is still large, so 
efforts are going on to find its evidence(s). If $R$-parity
is conserved, SUSY particles are pair produced, and the lightest neutralino 
in most of the cases is the LSP and thus dark matter candidate. Charginos
 ($\tilde \chi_{1,2,}^{\pm}$) and neutralinos ($\tilde \chi_{1,2,3,4}^{0}$) can decay into leptonic final states via superpartners of neutrinos ($\tilde \nu$, sneutrinos) or charged leptons 
($\tilde l$, sleptons), or via $W$, $Z$ or Higgs ($h$) bosons ($\tilde \chi_{i}^{\pm} \rightarrow \ell^{\pm}\tilde{\nu},
\nu{\tilde{\ell}}^{\pm},W^{\pm}\tilde\chi_{j}^{0},Z\tilde \chi_{j}^{\pm},h\tilde \chi_{j}^{\pm}$, and $\tilde \chi_{i}^{0} 
\rightarrow\nu\tilde \nu,\ell^{\pm}{\tilde{\ell}}^{\mp},W^{\pm}\tilde \chi_{j}^{\mp},Z\tilde\chi_{j}^{0},h\tilde\chi_{j}^{0}$, 
respectively). In recent studies the ATLAS and CMS Collaborations have reported new bounds on electroweakinos. For example
in Ref.~\cite{Aad:2014nua}, direct production of charginos and neutralinos is presented in events with three leptons 
and missing transverse 
energy $E_{T}^{miss}$ for 8 TeV center-of-mass energy. Here the simplified models are employed to study the direct pair 
production of
$\tilde \chi_{1}^{\pm}$ and $\tilde \chi_{2}^{0}$. $\tilde \chi_{1}^{\pm}$ and $\tilde \chi_{2}^{0}$ are assumed to be 
degenerate and consist purely of wino component. In this study, $\tilde \chi_{1}^{0}$ is assumed to be pure bino~\footnote{In the results given below, $m_{\tilde \chi_{1}^{0}}$=0 is assumed unless stated otherwise.}.
In such scenarios if $\tilde \chi_{1}^{\pm}$ and $\tilde \chi_{2}^{0}$ decay via the first-two generation sleptons and
sneutrinos $\tilde l/\tilde \nu$, 
their masses can be excluded up to 700 GeV. On the other hand if only $\tilde \tau/\tilde \nu_{\tau}$ as
the next to the LSP (NLSP) are involved while the first two generations of sleptons/sneutrinos are heavy,
 then the lower mass limit for 
$\tilde \chi_{1}^{\pm}$ and $\tilde \chi_{2}^{0}$ is 380 GeV. In case of $W/Z$ and $W/h$ mediated decays, 
$\tilde \chi_{1}^{\pm}$ and $\tilde \chi_{2}^{0}$ mass limits are 345 GeV and 148 GeV respectively. 
In another ATLAS SUSY searches~\cite{Aad:2014vma}, the direct productions of charginos, 
neutralinos, and sleptons in the final states with two leptons and missing transverse energy at 8 TeV 
center-of-mass energy is
reported. Here too, $\tilde \chi_{1}^{\pm}$ and $\tilde \chi_{2}^{0}$ are assumed to be degenerate and pure winos
while $\tilde \chi_{1}^{0}$ is pure bino.
In the scenario in which the masses of sleptons and sneutrinos lie between $\tilde \chi_{1}^{\pm}$ and $\tilde \chi_{1}^{0}$,
$\tilde \chi_{1}^{\pm}$ decays promptly to $l\nu\tilde \chi_{1}^{0}$ via $\tilde l\nu$ or $l^{\pm}\tilde \nu$, 
and its mass can be excluded in the range $[140,~465]$ GeV. On the other hand, if $\tilde \chi_{1}^{\pm}$ 
is the NLSP and decays via $W$ to $l\nu\tilde \chi_{1}^{0}$, its mass is excluded 
in the ranges $[100,~105]$ GeV, $[120,~135]$ GeV, and $[145,~160]$ GeV. In another scenario, $\tilde \chi_{1}^{\pm}$ and 
$\tilde \chi_{2}^{0}$ are considered mass degenerate and NLSPs, the direct $\tilde \chi_{1}^{\pm}\tilde \chi_{2}^{0}$ 
pair-production is followed by the decays $\tilde \chi_{1}^{\pm} \rightarrow W^{\pm}\tilde \chi_{1}^{0}$ and 
$\tilde \chi_{2}^{0}\rightarrow Z \tilde \chi_{1}^{0}$ with a 100$\%$ branching fraction. In this case the excluded mass range for
$\tilde \chi_{1}^{\pm}$ and $\tilde \chi_{2}^{0}$ is $[180,~335]$ GeV. A scenario in which slepton $\tilde l$ is the NLSP 
($pp\rightarrow \tilde l^{+}\tilde l^{-}\rightarrow l^{\pm} \tilde \chi_{1}^{0}$), the common values of the left- and right-handed
selectron and smuons masses between 90 GeV and 325 GeV are excluded, and for $m_{\tilde \chi_{1}^{0}}$=100 GeV the common 
values of the left- and right-handed slectron and smuons masses between 160 GeV and 310 GeV are excluded. Similar studies
have also been reported by the CMS Collaboration~\cite{CMS-PAS-SUS-13-006}. In the light of these
results, we investigate our data in Fig.~\ref{fig4}. The color coding for this figure is the following.
Blue points (black points in black and white print) represent the bino-type LSP 
neutralino, satisfy REWSB, and are consistent with the bounds on sparticle/Higgs masses including 
$123~{\rm GeV}\leqslant m_h\leqslant 127~{\rm GeV}$, B-physics, and
$4.7\times 10^{-10} \lesssim \Delta a_{\mu} \lesssim 52.7\times 10^{-10}.$ Red points (dark grey points in black and white print) are subset of blue points that
satisfy the WMAP9 5$\sigma$ bounds. The black solid lines are just to guide the eyes where we expect mass degeneracy in
LSP neutralino and other sparticle masses. We present plot in $m_{\tilde \chi_{1}^{\pm}}-m_{\tilde \chi_{1}^{0}}$ plane in 
the left top panel. Here blue and red points along the line represent the bino-wino coannihilation solutions. The chargino mass ranges 
for these points are $[140,~410]$ GeV and $\sim [180,~410]$ GeV for blue and red points, respectively. 
Interestingly, these red solutions are consistent with the bounds on the NLSP $\tilde \chi_{1}^{\pm}$ mentioned above. 
There is a horizontal strip of points along 
$m_{\tilde \chi_{1}^{0}}\approx$ 45 GeV, which are the $Z$-pole solutions. We also have another horizontal strip of red points
around $m_{\tilde \chi_{1}^{\pm}} \sim 430$ GeV and $m_{\tilde \chi_{1}^{0}} \sim 60$ GeV which represents Higgs-pole 
solutions. The above mentioned bounds on charginos do not apply on the resonance solutions. We will discuss 
these solutions later on. We also find blue and red points with $m_{\tilde \chi_{1}^{0}}\gtrsim$ 150 GeV 
and $100 \, {\rm GeV} \lesssim m_{\tilde \chi_{1}^{\pm}} \lesssim  700 \, {\rm GeV}$. These are the points where the 
NLSPs are sleptons or sneutrinos (either the {first two generations or third generation)}. We have
to be careful about all of these points and check their status. For this purpose,
 we display plots in $m_{\tilde \tau_1}-m_{\tilde \chi_{1}^{0}}$ and $m_{\tilde \nu_{\tau}}-m_{\tilde \chi_{1}^{0}}$ planes. 
In these plots, there are points where stau and tau sneutrino are the NLSPs and are degenerate in mass with the LSP 
neutralino. In order to make sure whether these
NLSP solutions satisfy the bounds on charginos discussed above, we present plots 
in $m_{\tilde \tau_1}/m_{\tilde \chi_{1}^{0}}-m_{\tilde \chi_{1}^{\pm}}$ 
and $m_{\tilde \nu_{\tau}}/m_{\tilde \chi_{1}^{0}}-m_{\tilde \chi_{1}^{\pm}}$ planes. These
plots clearly show that all the points $m_{\tilde \chi_{1}^{\pm}} \gtrsim$ 380 GeV are allowed as they satisfy the
chargino mass bounds in case of $\tilde \tau / \tilde \nu_{\tau}$-mediated chargino decays given above. In the
bottom left panel, we display plot in $m_{\tilde l}-m_{\tilde \chi_{1}^{0}}$ plane, which shows that 
most of our solutions easily satisfy
the upper bounds on the first-two generation slepton masses 325 GeV and 310 GeV 
respectively for $m_{\tilde \chi_{1}^{0}}$= 0 and 150 GeV.

\begin{figure}[htp!]
\centering
\subfiguretopcaptrue

\subfigure{
\includegraphics[totalheight=5.5cm,width=7.cm]{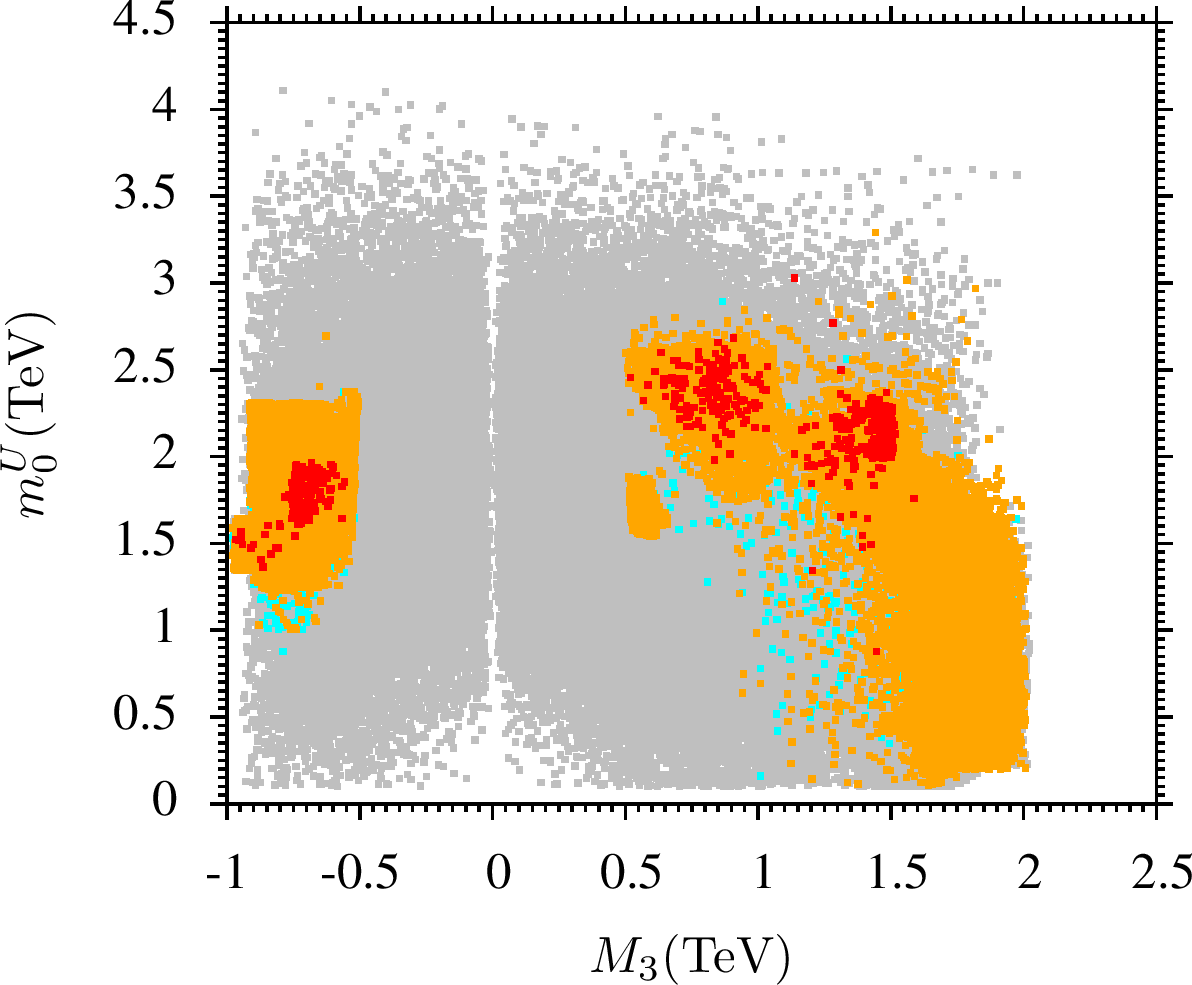}
}
\subfigure{
\includegraphics[totalheight=5.5cm,width=7.cm]{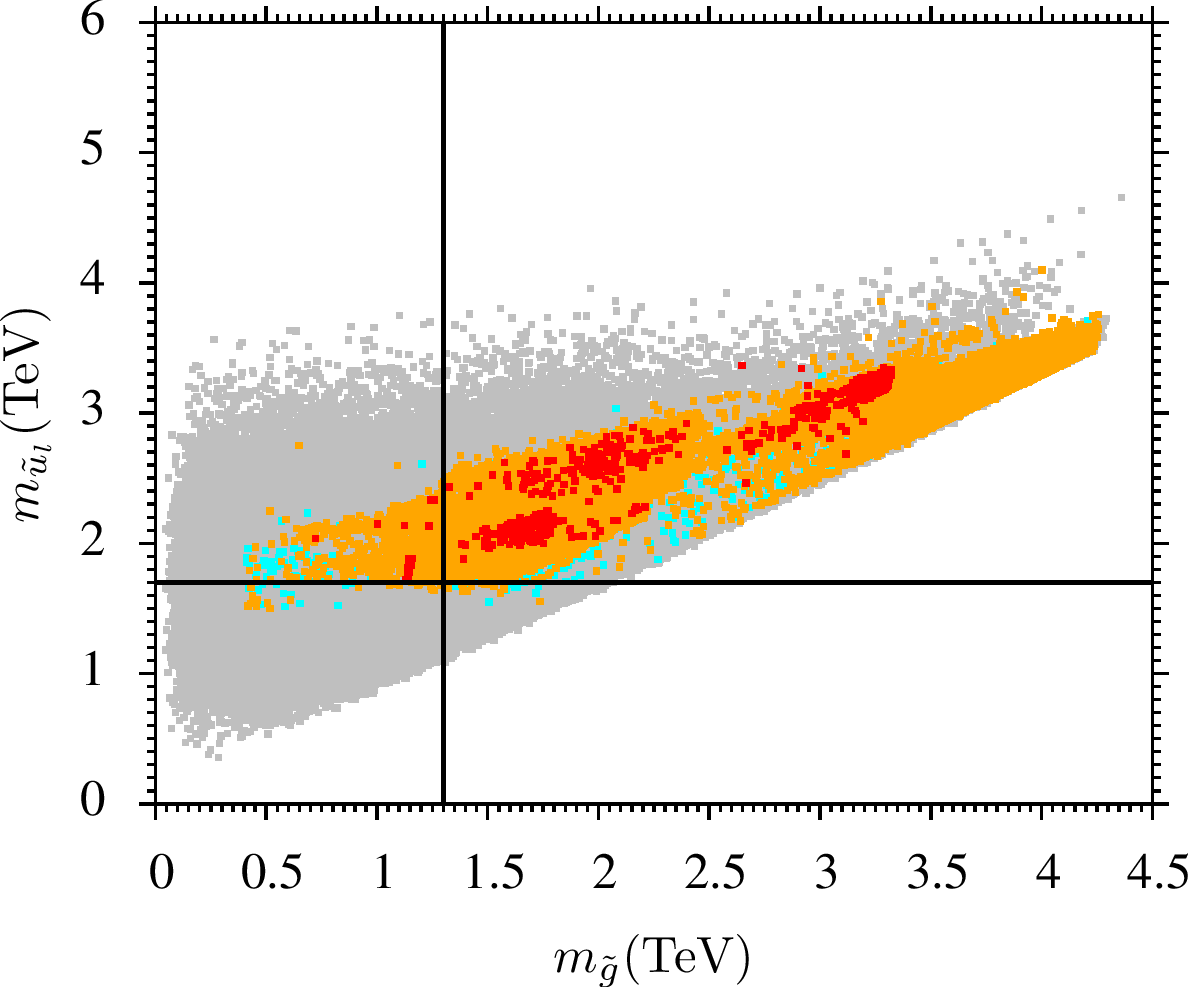}
}
\caption{Plots in $M_3-m_{0}^{U}$ and $m_{\tilde g}-m_{\tilde u_l}$ planes.
Grey points satisfy the REWSB and the lightest neutralino as an LSP conditions. Aqua points
(slightly dark grey points in black and white print) satisfy the sparticle mass bounds, B-physics bounds, and $123~{\rm GeV}\leqslant m_h\leqslant 127~{\rm GeV}$.
Orange points (slightly dark grey points in black and white print) form a subset of aqua points which satisfy the 3$\sigma$ bounds on $\Delta a_{\mu}$.
Red points (dark grey points in black and white print) are subset of orange points that also satisfy the WMAP9 5$\sigma$ bounds.
Also, we do not apply the squark and gluino mass bounds on the right panel, which are 
1.8~TeV and 1.3~TeV, respectively.}
\label{fig5_glu}
\end{figure}


In Fig.~\ref{fig5_glu}, we present plots in $M_3-m_{0}^{U}$ and $m_{\tilde g}-m_{\tilde u_l}$ planes.
Grey points satisfy the REWSB and neutralino as an LSP conditions. Aqua points
satisfy the mass bounds, B-physics bounds, and $123~{\rm GeV}\leqslant m_h\leqslant 127~{\rm GeV}$.
Orange points form a subset of aqua points that satisfy the 3$\sigma$ bounds on $\Delta a_{\mu}$.
Red points are subset of orange points which also satisfy WMAP9 5$\sigma$ bounds.
In the left panel orange points can be divided in three portions.
In case of large $M_3$ and large $m_{0}^{U}$, all colored sparticles are decoupled with masses around several TeV.
For large $M_3$ and small $m_{0}^{U}$, the colored sparticle spectra will be similar. 
However, when $M_3$ is small but $m_{0}^{U}$ is large, gluino is light around 
or below 1 TeV, while squarks are heavy. The mass squared (mass$^2$) of the right-handed squarks is predicted to be approximately 
twice of the left-handed ones. In the right panel, we show the mass ranges for gluino and left-handed squarks in our scans 
and do not apply squark and gluino mass bounds here. The black horizontal and vertical lines represent the squark and 
gluino bounds. For orange points, the gluino mass range is about $\sim [1300,4300]$ GeV corresponding to squark mass range 
 $\sim [1800,4000]$ GeV. While for red points, the upper limits on gluino and squark masses are relatively light 
about $ 3400$ GeV. Here we also note that because we have relatively light gluinos compared to $m_{\tilde g}\gtrsim 2$ TeV reported in Refs.~\cite{Cheng:2012np,Gogoladze:2014cha}, our parameter space can be probed easily 
at the next round of LHC supersymmetry searches.
It is shown in Ref.~\cite{cern_note1} that the squarks and gluino with masses around 
 2.5 TeV, 3 TeV, and 6 TeV may be probed by 
the LHC14, High Luminosity (HL) LHC14, and High Energy (HE) LHC33, respectively. This clearly shows that our models 
have testable predictions. Moreover, if we have collider facility with even higher energy in the future, 
we will be able to probe even larger values of sparticle masses.
\subsection{Dark Matter Relic Density}
\label{dm}

In this subsection we discuss the possible mechanism through which in our present scans, we get the observed dark
matter relic density, and also satisfy all the phenomenological bounds such as 
sparticle mass bounds, the Higgs boson mass bounds, 3$\sigma$ bounds on $\Delta a_{\mu}$, and 
B-physics bounds. We have already shown the existence of bino-wino coannihilation      
scenario in our model in the top left panel of Fig.~\ref{fig4}. Just to remind the reader, red points in that figure satisfy
 satisfy all the bounds just mentioned above.
We see red points along the black line with chargino mass 170-410 GeV.  Some portions of this mass range 
have already been explored by the LHC searches as discussed above. 
The International Linear Collider (ILC), a proposed $e^{+}e^{-}$ collider \cite{Baer:2013vqa,Baer:2013cma} 
was designed to operate at center-of-mass energy $\sqrt{s}\sim$ 0.25-1 TeV. At the ILC one 
can probe chargino mass up to $\sqrt{s}/2$. This clearly shows that the entire chrgino mass 
range mentioned above can be tested at the ILC, and we can have valuable information 
about SUSY contributions to $g-2$ indirectly.
Moreover, in the same plot we can see the $Z$-pole solutions 
with $m_{\tilde \chi_{1}^{0}} \sim$ 45 GeV. Such solutions are
constrained by the decay width of $Z$-boson to a pair of dark matter particles, $\Delta\Gamma (Z \rightarrow \chi_{1}^{0} \chi_{1}^{0})< 0.2$ 
~\cite{ALEPH:2005ab}. It was shown 
in Ref.~\cite{Han:2014nba} that this decay width
can be translated for bino-LSP case in terms of $\mu$ and $\mu \gtrsim$ 140 GeV is required 
in order to avoid experimental bound.
We have checked that all of our red points satisfy this bound. 

\begin{figure}[htp!]
\centering
\subfiguretopcaptrue

\subfigure{
\includegraphics[totalheight=5.5cm,width=7.cm]{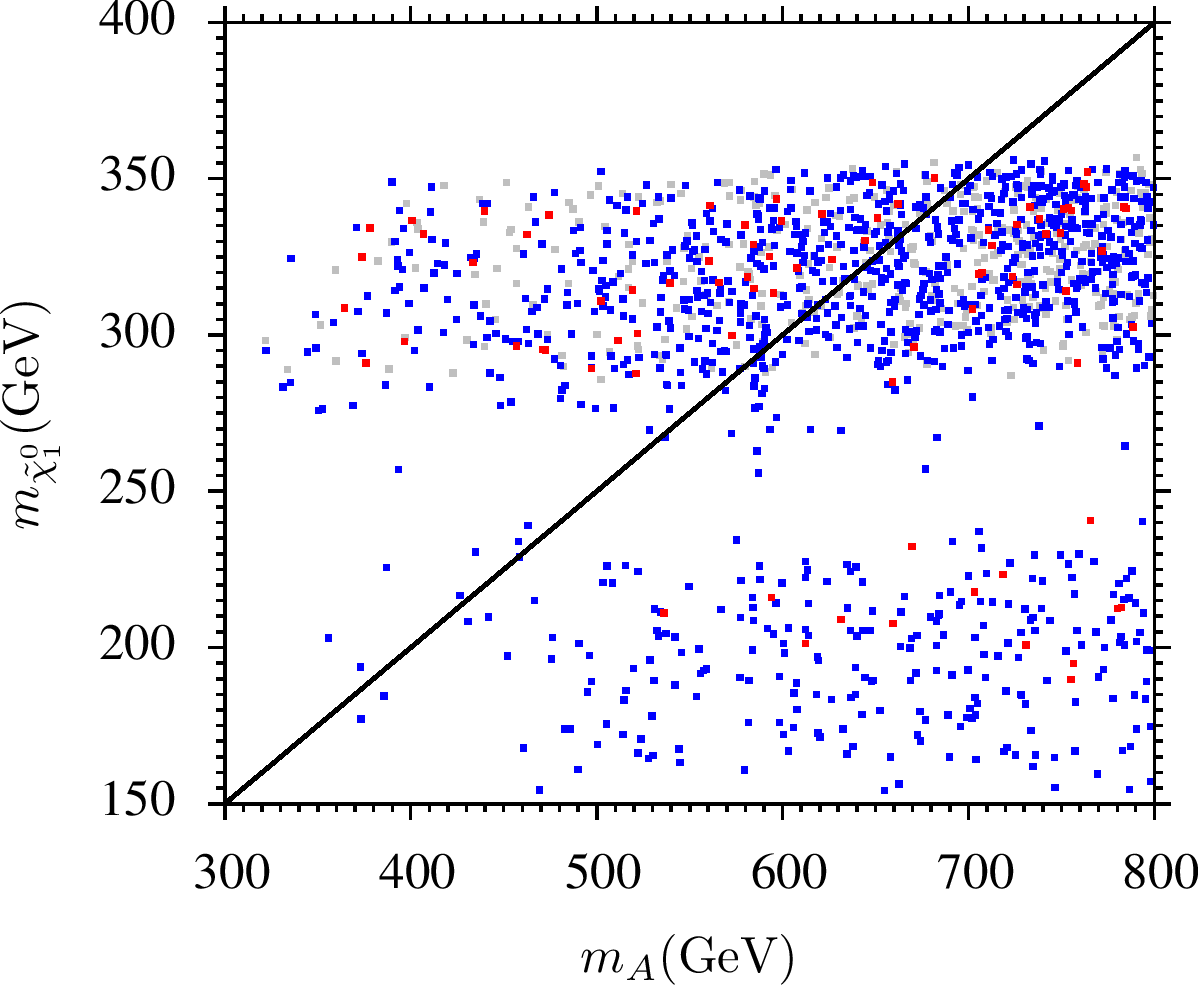}
}
\caption{Plot in $m_{\tilde \chi_{1}^{0}-m_A}$ plane.
The color coding is the same as in Fig.~\ref{fig4}.
}
\label{fig6_mA}
\end{figure}

\begin{figure}[htp!]
\centering
\subfiguretopcaptrue

\subfigure{
\includegraphics[totalheight=5.5cm,width=7.cm]{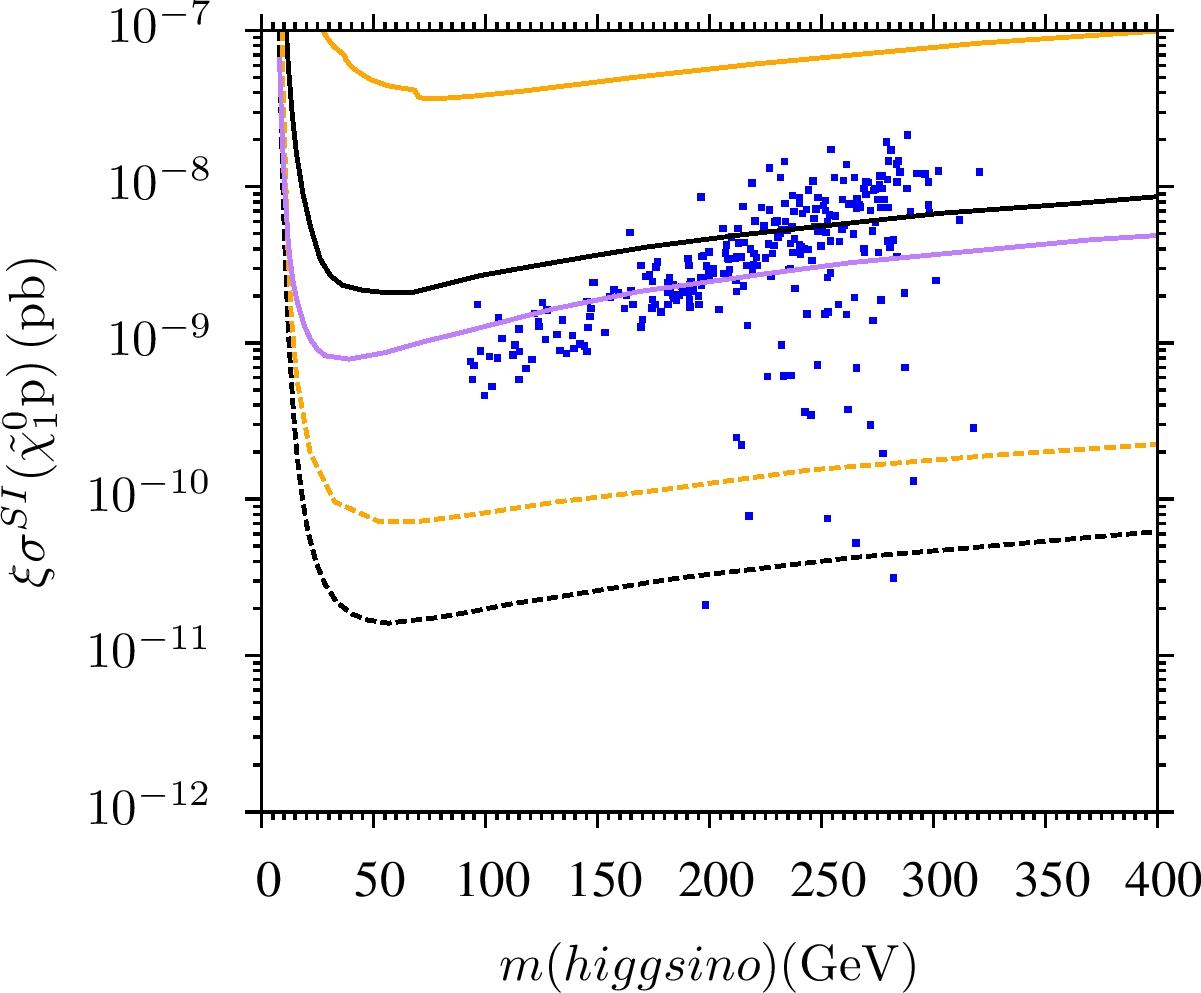}
}
\subfigure{
\includegraphics[totalheight=5.5cm,width=7.cm]{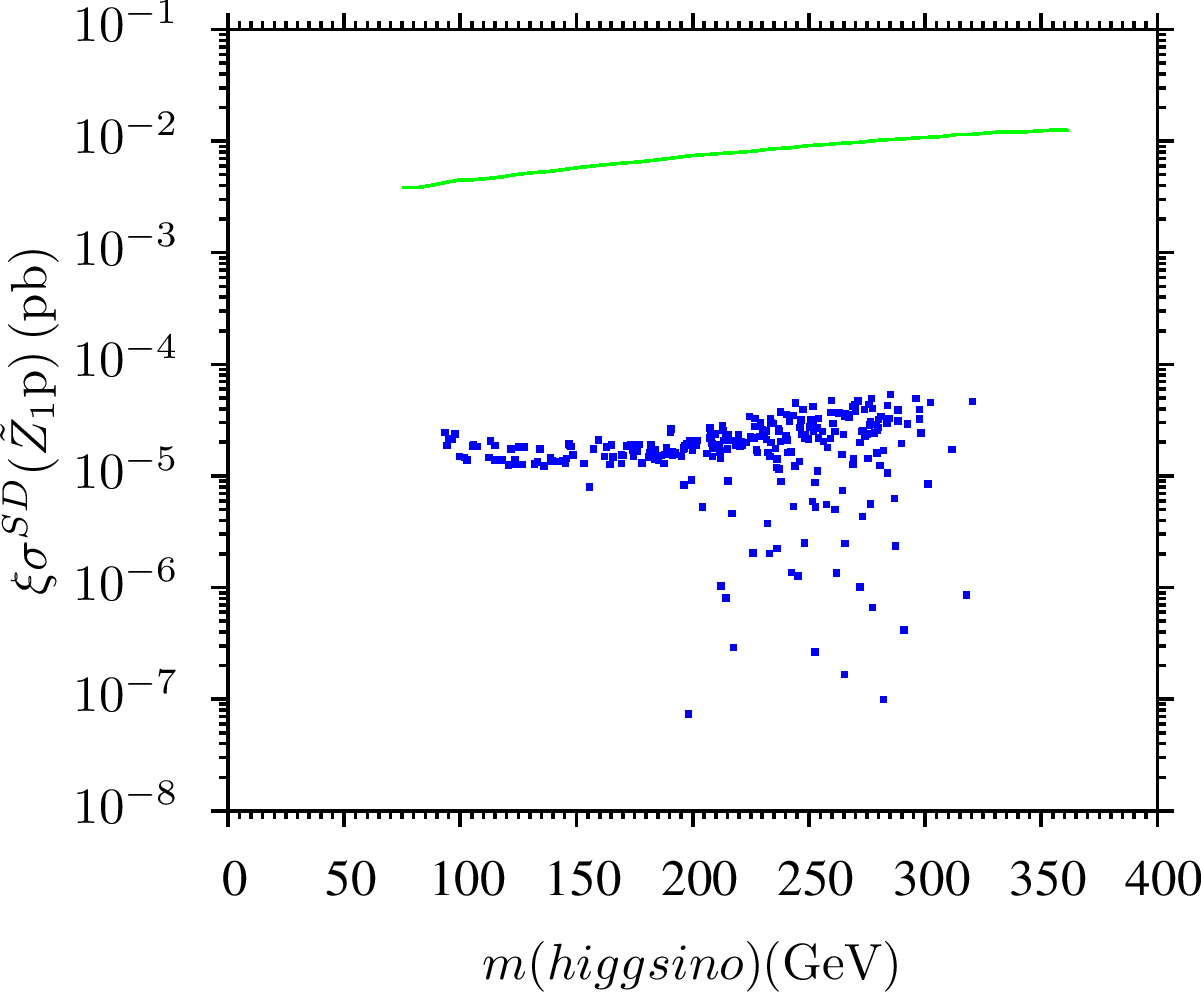}
}
\subfigure{
\includegraphics[totalheight=5.5cm,width=7.cm]{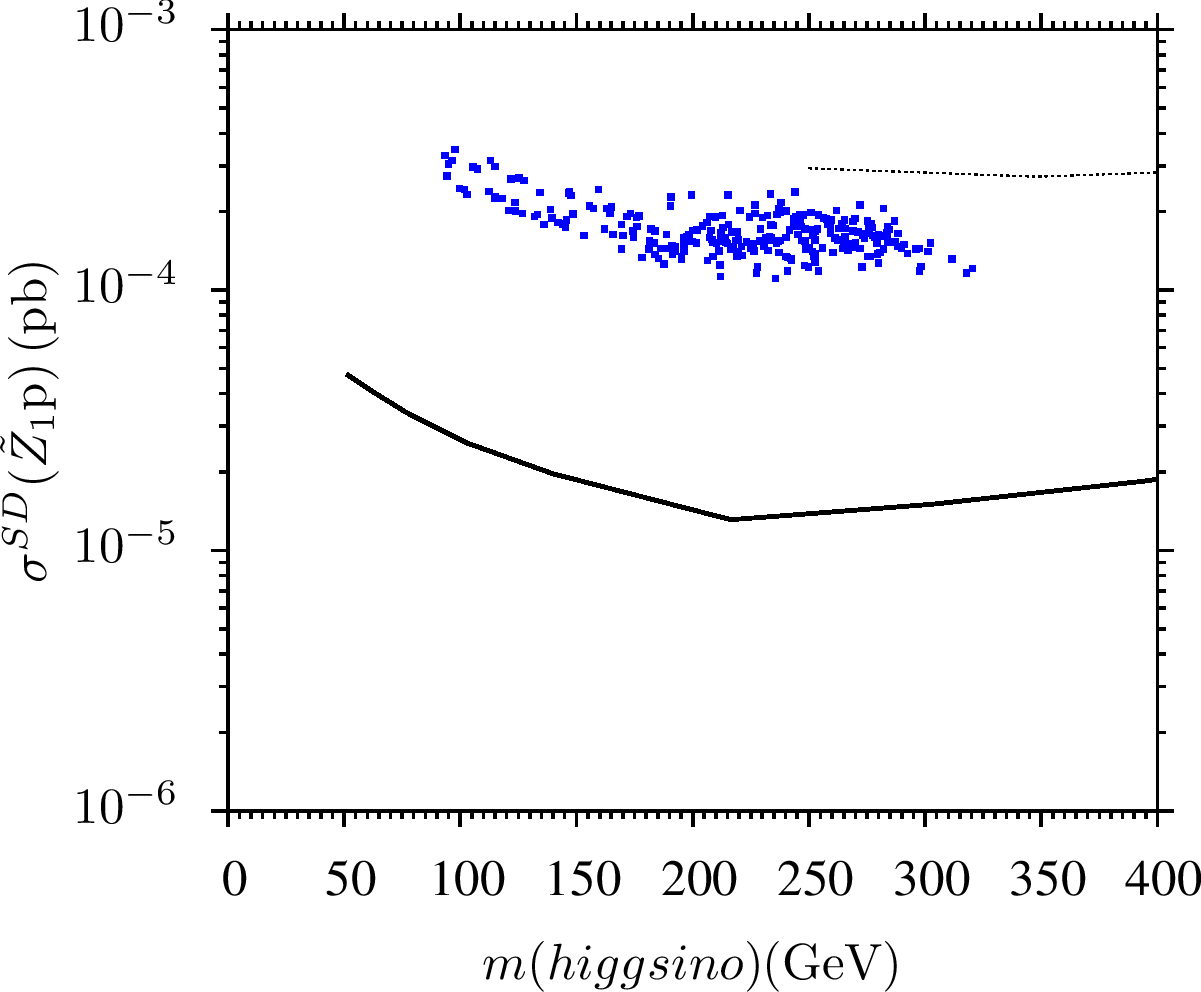}
}
\caption
{In the top left panel rescaled higgsino-like neutralino spin-independent cross section 
$\xi \sigma^{SI}(\tilde \chi^0_{1}p)$
versus $m({\rm higgsino})$ is shown. The orange solid line (top greyish solid line in black and white print) represents the current
 upper bound set by the CDMS experiment, black solid line depicts the upper bound set by XENON100 and the current upper bound set by the LUX experiment 
is shown by purple line (greyish solid line in black and white print), 
 while the orange (greyish in black and white print) and black dashed lines represent respectively the future reach
 of the SuperCDMS and XENON1T experiments. The top right panel shows plot in rescaled higgsino-like neutralino spin-dependent
cross section $\xi \sigma^{SD}(\tilde \chi^0_{1}p)-m(higgsino)$ plane. Green line (greyish solid line in black and white print) represents the upper bound set by the COUPP experiment. In the bottom panel (non-rescaled) higgsino-like neutralino spin-dependent cross section
$\sigma^{SD}(\tilde  \chi^0_{1}p)$ versus $m({\rm higgsino})$ 
is displayed. The IceCube DeepCore (black solid line) bound is
shown and the future IceCube DeepCore bound is depicted by the black dashed line.
Color coding is same as in Fig.~\ref{fig4}.
}
\label{fig7_xsection}
\end{figure}

In the same plane, there exist the Higgs-resonance solutions as a horizontal strip of red points around 
$m_{\tilde \chi_{1}^{0}}\sim 60$ GeV . We show Point 1 in Table \ref{table2} as an example of
such solutions. In particular, note that 
$Br(higgs\rightarrow {\tilde \chi_{1}^{0}}{\tilde \chi_{1}^{0}})\sim 4.68\times 10^{-4}$,  it is consistent with the 
results reported in Ref.~\cite{Belanger:2013kya}. In the top right and 
middle left panels of Fig.~\ref{fig4}, it is easy to see that we can accommodate the
LSP neutralino-stau and LSP neutralino-tau sneutrino coannihilation scenarios. The middle right and
bottom left panels of Fig.~\ref{fig4} show that in these scenarios, solutions with 
$\tilde \chi_{1}^{\pm}\gtrsim 380$ GeV do survive. Because the production cross-section of 
$\tilde \chi_{1}^{\pm} \tilde \chi_{2}^{0}$ is very 
large as compared to sleptons, it will be very hard to probe such solutions at the LHC. Apart from Higgs-pole and $Z$-pole
solutions we also have $A$-resonance solutions as can be seen in Fig.~\ref{fig6_mA}. The color coding is the same as in Fig.~\ref{fig4}, and the black line there represents
$m_{A}=2m_{\tilde \chi_{1}^{0}}$. For blue points, we have $m_A$ as light as 370 GeV and as heavy as
700 GeV along the line. While for red points, the lower limit for $m_A$ is about 600 GeV. Apart from our red points
that satisfy the 5$\sigma$ WMAP9 bound in Figs.~\ref{fig4} and \ref{fig6_mA}, we would like to comment on blue points. 
These blue solutions have $\Omega h^2$ values either above or below the 5$\sigma$ WMAP9
bounds. To solve this problem, in the former case, one can treat bino as the NLSP and assume 
that it decays to a lighter state, 
for example, $\tilde \chi_{1}^{0}\rightarrow \gamma \tilde a$ where $\tilde a$ is axino. In such a scenario we will
have the mixed axion and axino ($a\tilde a$) dark matter~\cite{pqww}. In the latter case where we have relic density 
$\Omega h^{2} \sim ~ 10^{-5}-10^{-2}$, the neutralino abundance can be accommodate in the Pecci-Quinn 
augmented MSSM, where $m_{\tilde a} > m_{\tilde \chi_{1}^{0}}$ and additional neutralinos are produced via thermal axino
production and decay $m_{\tilde a} \rightarrow m_{\tilde \chi_{1}^{0}} \gamma$~\cite{bls}. In these cases, the cold dark
matter tends to be neutralino dominant with a small component of axions.
In addition to the bino-type neutralino LSP, we  have the wino-type and higgsino-type  neutralino LSPs
as well. Let us discuss them one by one.
It was shown in Refs.~\cite{Fan:2013faa, Cohen:2013ama} that 
for Navarro-Frenk-White and Einasto distributions, the entire mass range of thermal wino dark matter 
from 0.1 to 3 TeV may be excluded.
In a recent study \cite{Hryczuk:2014hpa}, it was shown that wino as dark matter candidate is excluded 
in the mass range below 800 GeV from antiproton and between 1.8 TeV to 3.5 TeV from 
the absence of a $\gamma$-ray line feature 
toward the galactic center. Because our wino-type solutions have very small relic density
from $10^{-3}$ to $10^{-5}$, for example, Point 5 in Table~\ref{table1},
the light wino like LSP neutralino, which can provide a solution to the
$a_{\mu}$ anomaly, does satisfy the above constraints. 
Even if one has a thermal wino-like LSP neutralino with mass around 2.8 TeV and
the observed relic density, one can escape the above bounds by assuming that the wino-like neutralino is 
the NLSP and decays to axino and $\gamma$.
Another example of solutions with under abundance relic density is the higgsino-like LSP.  In order to 
match the observed dark matter relic density, we need an additional dark matter candidate 
along with higgsino. In this scenario the higgsino could make only a small fraction of the dark matter relic density
 and the remaining abundance is composed of axinos produced through the vacuum misalignment~\cite{vacmis}. This also provides
the possibility to detect axinos along with the chances to detect higgsinos despite the fact that their relic
density is somewhat suppressed between 1-15 in the present Universe. In the top left panel of Fig.~\ref{fig7_xsection} 
we plot the rescaled higgsino-like neutralino spin-independent cross section 
$\xi \sigma^{SI}(\tilde \chi^0_{1}p)$ versus $m({\rm higgsino})$.
The orange solid line (top greyish solid line in black and white print) represents the current  upper bound set by the CDMS experiment, 
the black solid line depicts the upper bound set by the XENON100 experiment~\cite{Xenon100},
 and the current upper bound set by the LUX experiment
~\cite{Akerib:2013tjd} is shown by purple line (greyish solid line in black and white print), while the orange (greyish in black and white print) and black dashed lines represent respectively the future reach
 of the SuperCDMS~\cite{Brink:2005ej} and XENON1T \cite{Aprile:2012zx} experiments.
In order to account for the
fact that the local higgsino relic density might be much less than the usually assumed value
 $\rho_{local}\simeq 0.3 \,{\rm GeV}/cm^{3}$,
we rescale our results by a factor $\xi=\Omega_{\tilde \chi^0_1}h^{2}/0.11$~\cite{Bottino:2000jx}. Blue points 
satisfy all the bounds mentioned in
Section~\ref{sec:scan} except the WMAP9 bounds. 
We note that these solutions have $50 \lesssim \Delta_{EW} \lesssim 130$ and most
of the solutions have $\Delta_{EW}\lesssim 100$. 
However, the solutions
 with $m({\rm higgsino})$ in the range $[100,~325]$ GeV
can be ruled out by the LUX experiment depending upon their $\xi \sigma^{SI}(\tilde \chi^0_{1}p)$ values. The
 rest of the solutions with small values of $\xi \sigma^{SI}(\tilde \chi^0_{1}p)$ will be probed by the SuperCDMS and XENON1T experiments but 
not completely.
Here we would like to comment on our solutions just below the XENON1T reach 
line with higgsino mass around $m({\rm higgsino})\sim 200$ GeV and 300 GeV 
which have $\mu \sim$ 209 GeV and 313 GeV, while $\Delta_{EW}\sim$ 102 and 128, respectively. The presence of this point shows that it would be difficult
to rule out the higgsino-like LSP neutralino for 
entire parameter space in $R$-parity conserving natural SUSY models~\cite{Baer:2013vpa}.
The top right panel shows a plot in the rescaled higgsino-like neutralino spin-dependent
cross section $\xi \sigma^{SD}(\tilde \chi^0_{1}p)$ as a function of $m({\rm higgsino})$. Green line represents 
the upper bound set by COUPP experiment~\cite{Behnke:2012ys}.  We see that our solutions 
are about a couple of magnitude below the current bounds from the COUPP experiment. For comparison, 
in the bottom panel of Fig.~\ref{fig7_xsection}, we present a plot of the (non-rescaled) 
higgsino-like neutralino spin-dependent cross section $\sigma^{SD}(\tilde \chi^0_{1}p)$ versus
 $m({\rm higgsino})$. The IceCube DeepCore and future IceCube DeepCore 
bounds are shown in black solid line and black dashed line~\cite{IceCube}.
The color coding is the same as the left panel. Because the IceCube detection 
depends on whether the Sun has equilibrated its core abundance between capture rate 
and annihilation rate~\cite{Jungman:1995df}, we do not rescale our results here.
It was shown in~\cite{Niro:2009mw} that for the Sun, equilibration is reached 
for almost all of SUSY parameter space. 
If this is true, then our solutions will be probed by the future IceCube DeepCore experiment.
However, we are not sure whether such equilibration can be reached if the SUSY
particles are relatively heavy.

\begin{table}[b]\hspace{-1.0cm}\vspace{-0.3cm}
\centering
\begin{tabular}{|c|ccccc|}
\hline
\hline

                 & Point 1 & Point 2 & Point 3 & Point 4 & Point 5\\

\hline
$m_{\tilde Q}$      &1268.4  & 1435.2   &2141   & 2095.2  & 467.95\\
$m_{\tilde U^c}$    &1632.2  & 1914.5   &3022.7 & 2872.3  & 161.02 \\
$m_{\tilde D^c}$    &1736.7  & 1988.1   &3004.9 & 2899.4  & 349.3 \\
$m_{\tilde L}$      &159.6   & 152.9    &438.2  & 543.7   & 517.6 \\
$m_{\tilde E^c}$    &743.9   & 673.7    &175.6  & 728.1   & 641.9   \\
$M_{1} $            &108.9   & 112.1    &515    & 899.8   & 786.1\\
$M_{2}$             &706.1   & 682.4    &287.1  & 495.6   & 158.5 \\
$M_{3}$             &-786.9  & -743.35  &856.85 & 1506.1  & 1727.5\\
$A_t=A_b$           &3564    &3725     &-5073  & -5897    & -4817 \\
$A_{\tilde \tau}$   &-496.7  &-465.7   &168.7  &  396.1  &  -250.6\\
$\tan\beta$         &15.9     &17.9     &25.9   & 28.6    & 31.2\\
$m_{H_u}$           &2457     &2581     &3160   & 2306    & 1856\\
$m_{H_d}$           &2507     &2523     &562.6  & 265.9   & 288.8\\
\hline
$\mu$               &162      &207     & 1572   & 3070 & 2553\\
$\Delta_{EW}$       &16.5      &25     & 598    &2269  & 1569\\
$\Delta_{HS}$       &1462      &1616     & 3005     &3540  & 2393\\
$\Delta a_{\mu}$    &$11.62 \times 10^{-10}$ &$15.12\times 10^{-10}$ &$22.40\times 10^{-10}$  &$7.78\times 10^{-10}$ &$11.67\times 10^{-10}$ \\
\hline

\hline
$m_h$            &123  & 123 & 124  & 125  & 125\\
$m_H$            &2446 &2407 & 708  & 2524 & 2081\\
$m_A$            &2430 &2391 & 703  & 2507 & 2068\\
$m_{H^{\pm}}$    &2447 &2408 & 713  & 2525 & 2083\\
\hline
$m_{\tilde{\chi}^0_{1,2}}$
                 &46, 165   &48, 208   &218, 234      &387, 404  &108, 329  \\

$m_{\tilde{\chi}^0_{3,4}}$
                 &173, 611   &218, 591   &1571, 1574 &3053, 3054  &2544, 2544 \\

$m_{\tilde{\chi}^{\pm}_{1,2}}$
                 &165, 605   &211, 584  &234, 1570   &406,3059  &108, 2547 \\
\hline
$m_{\tilde{g}}$  &1829   &1752  & 2029  & 3323 & 3672\\
\hline
$m_{ \tilde{u}_{L,R}}$
                 &2036, 2233   &2081, 2382 & 2689, 3411 &3491, 3986  &3211, 3227 \\
$m_{\tilde{t}_{1,2}}$
                 &830, 1401  &966, 1387    &1444, 1867  &2298, 2592  &1862, 2515  \\
\hline
$m_{ \tilde{d}_{L,R}}$
                 &2038, 2322   &2083, 2469    &2690, 3433 &3492, 4036 &3212, 3203  \\
$m_{\tilde{b}_{1,2}}$
                 &1384, 2217   &1354, 2342    &1533, 3176 &2486, 3723  &2492, 2894  \\
\hline
$m_{\tilde{\nu}_{1,2}}$
                 &442  &367  & 267  & 497 & 621\\
$m_{\tilde{\nu}_{3}}$
                 & 365  &240  &287  & 502 & 622\\
\hline
$m_{ \tilde{e}_{L,R}}$
                &457, 767   &384, 755   &251, 552   &504, 943  & 644, 501  \\
$m_{\tilde{\tau}_{1,2}}$
                & 383, 691    &266, 654   &247, 571 &457, 948  & 402, 685 \\
\hline

$\sigma_{SI}({\rm pb})$
                & $1.54\times 10^{-9}$ & $ 7.14\times 10^{-10} $ & $ 4.9\times 10^{-11} $ & $8.09\times 10^{-13}$ & $4.65\times 10^{-12}$\\

$\sigma_{SD}({\rm pb})$
                & $2.13\times 10^{-4}$ &$ 7.99\times 10^{-5} $ & $ 2.62\times 10^{-8} $ & $7.91\times 10^{-10}$ & $3.09\times 10^{-8}$\\

$\Omega_{CDM}h^{2}$&  0.006 &0.122  &0.096  &0.103  &0.0007 \\
\hline
\hline
\end{tabular}
\caption{All the masses in this table are in units of GeV.
All the points satisfy the constraints  described in 
Section~\ref{sec:scan}. Points 1 and 2 display the solutions with the minimal values of $\Delta_{EW}$ which are respectively 
 not consistent and consistent with the $5\sigma$ WMAP9 bounds. Point 2 is an example of $Z$-pole solutions. 
Point 3 represents a solution with large contribution to $\Delta a_{\mu}$ and consistent with the $5\sigma$ WMAP9 bounds. 
Point 4 and 5 are the examples with large gluino masses, 125 GeV light CP-even Higgs boson mass, which are
respectively consistent and not consistent with the $5\sigma$ WMAP9 bounds. Points 3 and 4 are 
 the bino-wino coannihilation scenario while Point 5 represents wino-like LSP solutions. 
}
\label{table1}
\end{table}
\begin{table}[b]\hspace{-1.0cm}
\centering
\begin{tabular}{|c|ccccc|}
\hline
\hline
               & Point 1    & Point 2 & Point 3 & Point 4 & Point 5\\

\hline
$m_{\tilde Q}$    &1631.4  &2305.6 & 2328.6   & 2344  & 2084.3  \\
$m_{\tilde U^c}$  &2302.7  &3230.9 & 3258.8   & 3288.9  & 2907.8  \\
$m_{\tilde D^c}$  &2274  &3246.1 & 3277.2     & 3301.6  & 2928   \\
$m_{\tilde L}$    &466.8  &211.7   &210.2    & 215.1   & 237.8   \\
$m_{\tilde E^c}$  &143.6  &439.5   &473.7     & 414   & 483.4      \\
$M_{1} $          &135.4  &709.5   &768.2    & 761.2   & 789.7     \\
$M_{2}$           &658.8  &687.4   &742.5    & 701.0   & 707.2   \\
$M_{3}$           &-649.7  &742.65  &806.75   & 851.5  & 913.45  \\
$A_t=A_b$         &4095  &-4616   &-4675    & -4695   & -4078   \\
$A_{\tilde \tau}$ &-202.5  &918     &807.3    & 846.7   & 784.8  \\
$\tan\beta$       &21.5  &14.2    &15.9     & 14.7    & 16.1    \\
$m_{H_u}$         &2708  &3641    &3295     & 3521    & 3472    \\
$m_{H_d}$         &2722  &828.3   &954.6    & 980.1   & 861  \\
\hline
$\mu$             &451  &503  & 1459     & 1170    & 209   \\
$\Delta_{EW}$     &49  &76   & 512     &  329   & 102  \\

$\Delta_{HS}$     &1818  &3260   & 3128     &  3318   & 2914  \\

$\Delta a_{\mu}$  & $12.2\times 10^{-10}$  &$7.5 \times 10^{-10}$ &$5.76\times 10^{-10}$ &$6.0\times 10^{-10}$  &$10.3\times 10^{-10}$  \\
\hline

\hline
$m_h$          &123  &123  & 123 & 123  & 123\\
$m_H$          &2503  &577  & 1499 & 1275 & 401\\
$m_A$          &2487  &573  & 1490 & 1267 & 398\\
$m_{H^{\pm}}$  &2504  &582  & 1502 & 1278 & 409 \\

\hline
$m_{\tilde{\chi}^0_{1,2}}$
               &60, 433  &300, 478   & 331, 611   &327, 573 &198, 219 \\

$m_{\tilde{\chi}^0_{3,4}}$
               &462, 586  &513, 604  &1466, 1469   & 1179, 1184 &344, 589 \\
$m_{\tilde{\chi}^{\pm}_{1,2}}$
               &440, 580  &481, 598   &611, 1469    &573, 1184 & 214, 578\\
\hline
$m_{\tilde{g}}$ &1571  &1798   & 1933  & 2025 & 2132\\
\hline 
$m_{ \tilde{u}_{L,R}}$
               &2113, 2567  &2715, 3515    &2805, 3576    &2857, 3642 & 2725, 3383 \\
$m_{\tilde{t}_{1,2}}$
               &1034,1323  &1544, 1868   &1726, 2062     &1741, 2064 & 1648, 1914 \\
\hline 
$m_{ \tilde{d}_{L,R}}$
              & 2114, 2632  &2716, 3544    &2806, 3625     & 2858, 3608 & 2726, 3401 \\
$m_{\tilde{b}_{1,2}}$
               &1222, 2444  &1631, 3453    &1813, 3516     & 1832, 3582 & 1766, 3295 \\
\hline
$m_{\tilde{\nu}_{1,2}}$
               &378  &400  & 358  &350 & 489\\
$m_{\tilde{\nu}_{3}}$
              & 166  & 385  & 336  &330 &474\\
\hline
$m_{ \tilde{e}_{L,R}}$
              &387, 689  &399, 582     &352, 731   &345, 662 & 493, 564\\
$m_{\tilde{\tau}_{1,2}}$
              &184, 498  &395, 582     &340, 715    &337, 646 & 485, 546 \\
\hline

$\sigma_{SI}({\rm pb})$
              &$7.42 \times 10^{-11}$  & $2.7 \times 10^{-9}$ & $ 1.7 \times 10^{-11} $ & $ 4.11\times 10^{-11} $ & $ 4.66\times 10^{-8} $\\

$\sigma_{SD}({\rm pb})$
             &$3.44\times 10^{-6}$   & $5.82\times 10^{-6}$ &$ 2.82\times 10^{-8} $ & $ 8.06\times 10^{-8} $  & $ 1.64 \times 10^{-4} $\\

$\Omega_{CDM}h^{2}$&0.129 & 0.098  & 0.124  &0.123  & $4.94\times 10^{-5}$\\
\hline
\hline
\end{tabular}
\caption{All the masses in this table are in units of GeV.
All  points satisfy  all the constraints in Section~\ref{sec:scan} except point 4 which does not satisfy the $5\sigma$
WMAP9 bounds. Points 1, 2, 3, and 4 represent the Higgs-resonance, $A$-resonance, neutralino-stau coannihilation, 
and neutralino-stau neutrino  coannihilation solutions, respectively.
Point 5 is an example of the higgsino-like LSP. This point has the 
rescaled higgsino-like neutralino spin-independent cross section $\xi \sigma^{SI}(\tilde Z_{1}p)$ 
below the XENON1T experimental
upper bound ($\Omega_{\tilde \chi^0_1}h^{2}/0.11 \times \sigma^{SI}(\tilde \chi^0_{1}p) \sim 2.09\times 10^{-11}{\rm pb}$).   
}
\label{table2}
\end{table}

In Table~\ref{table1}, we present five benchmark points. All the points satisfy the bounds 
on the sparticle and Higgs boson masses as well
as the constraints from B-physics and $\Delta a_{\mu}$  described in 
Section~\ref{sec:scan}. Points 1 and 2 are the solutions with the minimal values 
of $\Delta_{EW}$ that are respectively not consistent and consistent with the WMAP9 5$\sigma$ bound. 
Here we see that the mass of the bino-like LSP neutralino is about 
$\sim$ 46-48 GeV, $m_{\tilde \chi_{1}^{\pm}}$ range is $\sim [165,~211]$ GeV while 
the CP-even Higgs boson mass is around 123 GeV. For the first 
two-family sleptons and sneutrinos, the left-handed sleptons are
lighter than the right-handed sleptons and are in the mass range $[380,~460]$ GeV, 
while the third-family light stau and tau sneutrino 
can be as light as 266 GeV and 240 GeV, respectively. For the colored sparticles, gluino mass is around 1800 GeV, 
the first-two family squarks are in the mass range $\sim [2000,2400]$ GeV, and light stop is around 830 GeV. 
Point 2 also represents the $Z$-pole solutions. Point 3 represents a solution with large 
contribution to $\Delta a_{\mu} \sim 22.4 \times 10^{-10}$ (within one $\sigma$ bound on $\Delta a_{\mu}$) and 
consistent with the $5\sigma$ WMAP9 bounds. It is also an example of the bino-wino
connihilation scenario with $m_{\tilde \chi_{1}^{0}}\sim$ 218 GeV and $m_{\tilde \chi_{1}^{\pm}}\sim$ 234 GeV. 
Here, we see that the Higgs boson mass is around 124 GeV, 
the first two families of left- and right-handed sleptons respectively have masses $\sim$ 251 GeV and
552 GeV, the light stau mass is around 247 GeV, and the tau sneutrino mass is 287 GeV. Gluino mass is around 2000 GeV, 
while the first two families of squark masses are from 2680 to 3430 GeV. The light stop mass is around 1444 GeV.
Points 4 and 5 respectively are the examples of solutions with large gluino masses about 3323 GeV and 4215 GeV and 
125 GeV light CP-even Higgs boson mass. Point 4 is consistent with relic density bounds with 
$\Delta a_{\mu}\sim7.87 \times 10^{-10}$, while Point 5 have $\Delta a_{\mu}\sim 11.54 \times 10^{-10}$ but do not satisfy 
relic density bound. Point 4 is another example of the bino-wino coannihilation scenario.
Point 5 is the representative of the wino-like LSP neutralino solutions. For Points 4 and 5,
the first two families of right-handed sleptons are respectively 943~GeV and 870~GeV,
but the corresponding left-handed sleptons are 504~GeV and 477~GeV. Also,
sneutrinos have masses 497 GeV and 430 GeV, respectively. The light stau and tau sneutrino masses for 
Point 4 are 457 GeV and 502 GeV while for
Point 5 light stau mass is 136 GeV and tau sneutrino mass is about 283 GeV.

In Table~\ref{table2}, we display another five benchmark points consistent with the constraints described in Section~\ref{sec:scan}. 
Points 1, 2, 3, and 4 represent Higgs-resonance, $A$-resonance, neutralino-stau and neutralino-tau 
sneutrino solutions, respectively. Point 5 is an example of the higgsino-like LSP. 
Here we see that all of these points have a lot of common features. Gluino masses are in the range $[1700,~2100]$ GeV while
the first two families of squarks have masses from around 2700 GeV to 3600 GeV. The light stop mass lies in 
the range $[1000,~1750]$ GeV.
Also, the first two families of sleptons and sneutrinos are almost degenerate. In both tables
of benchmark points the light stop is the lightest colored sparticle. Point 5 has the rescaled higgsino-like neutralino 
spin-independent cross section $\xi \sigma^{SI}(\tilde \chi^0_{1}p)=\Omega_{\tilde \chi^0_1}h^{2}/0.11 \times 
\sigma^{SI}(\tilde \chi^0_{1}p) \sim 2.09\times 10^{-11}{\rm pb}$, which is below the XENON1T experimental low bound.

\section{Discussions and Conclusion}
\label{summary}

We attempted to resolve the muon $(g-2)_{\mu}/2$ discrepancy in the SM by exploring the MSSM with the EWSUSY from GmSUGRA.
We identified a viable parameter space that resolves this discrepancy, and as a by product we obtained the solutions
with small EWFT simultaneously. Our solutions not only provide sizable contributions to $\Delta a_{\mu}$ but also
satisfy all the current experimental constraints including the LHC SUSY searches. In particular,
the relic density for cold dark matter can be achieved within the 5$\sigma$ WMAP9 bounds by the bin-wino, neutralino-stau, 
neutralino-tau sneutrino coannihilation scenarios, and the $A$, Higgs and $Z$ resonance scenarios. 
Moreover, we identified the higgsino-like LSP neutralino and 
calculate the spin-independent and spin-dependent cross sections on the LSP neutralinos with nucleons. 
\section*{Acknowledgements}
We would like to thank Howard Baer for useful discussions. This research was supported in part by the Natural Science
Foundation of China under grant numbers 10821504, 11075194, 11135003, 11275246, and 11475238,
 and by the National
Basic Research Program of China (973 Program) under grant number 2010CB833000 (TL).


\end{document}